\documentstyle[12pt]{article}

\begin{document}


\hsize=6.15in
\vsize=8.2in
\hoffset=-0.42in
\voffset=-0.3435in

\normalbaselineskip=24pt\normalbaselines

"Journal of Computational Neuroscience", in press.

\vspace{1.75cm}

\begin{center}
{\large \bf Thermodynamic constraints on neural dimensions, firing rates,
brain temperature and size}
\end{center}

\vspace{1.15cm}

\begin{center}
{Jan Karbowski}
\end{center}

\vspace{2.05cm}

\begin{center}
{\it Sloan-Swartz Center for Theoretical Neurobiology,
Division of Biology 216-76, \\
California Institute of Technology,
Pasadena, CA 91125, USA; \\ 
Institute of Biocybernetics and Biomedical Engineering, \\
Polish Academy of Sciences, 02-109 Warsaw, Poland }
\end{center}


\vspace{2.1cm}

\noindent Email: jkarb@its.caltech.edu

\newpage

\begin{abstract}
There have been suggestions that heat caused by cerebral metabolic 
activity may constrain mammalian brain evolution, architecture, and 
function. This article investigates physical limits on brain 
wiring and corresponding changes in brain temperature that are 
imposed by thermodynamics of heat balance determined mainly by 
Na$^{+}$/K$^{+}$-ATPase, cerebral blood flow, and heat conduction. 
It is found that even moderate firing rates cause significant
intracellular Na$^{+}$ build-up, and the ATP consumption rate
associated with pumping out these ions grows nonlinearly with
frequency. Surprisingly, the power dissipated by the Na$^{+}$/K$^{+}$ 
pump depends biphasically on frequency, which can lead to the biphasic
dependence of brain temperature on frequency as well. Both the total 
power of sodium pumps and brain temperature diverge for very small 
fiber diameters, indicating that too thin fibers are not beneficial for
thermal balance. For very small brains blood flow is not a sufficient
cooling mechanism deep in the brain. The theoretical lower bound on 
fiber diameter above which brain temperature is in the operational regime 
is strongly frequency dependent but finite due to synaptic depression.
For normal neurophysiological conditions this bound is at least an order 
of magnitude smaller than average values of empirical fiber diameters, 
suggesting that neuroanatomy of the mammalian brains operates in the 
thermodynamically safe regime. 
Analytical formulas presented can be used to estimate average 
firing rates in mammals, and relate their changes to changes in brain 
temperature, which can have important practical applications. 
In general, activity in larger brains is found to be slower than
in smaller brains.

\end{abstract}





\vspace{0.1cm}



\vspace{0.3cm}

\newpage

\section{Introduction}

Brain like any computational device (Landauer 1961; Bennett, 1982)
or biological organ (Rolfe and Brown, 1997) dissipates energy 
(Siesjo, 1978; Ames, 2000; Clarke and Sokoloff, 1994). 
Recent studies on cerebral metabolism indicate that brain
is energy expensive (Aiello and Wheeler, 1995; Attwell and Laughlin,
2001; Lennie, 2003), as it consumes relatively more energy than most 
other organs, and its total metabolic rate has larger scaling exponent 
than whole body metabolic rate (Karbowski, 2007). Potential thermal 
imbalance between heat produced and removed could lead to overheating 
and aberrant functioning, because neural properties are strongly 
temperature dependent (Koch, 1998). Therefore, the problem of heat 
transfer could be a serious factor shaping brain evolution (Falk, 1990), 
organization, and function (Baker, 1982; Raichle, 2003; Kiyatkin, 2007).
Consequently, what is the relationship between brain temperature, its
size, cerebral power generated and neural activity? Can we estimate 
changes in firing rates based
on changes in temperature? Is there any thermodynamic limit on brain 
size? If so, does 5 kg, which is the mass of the largest mammalian brain 
(Haug, 1987; Purves, 1988), approach that limit? Are neural sizes
and fibers also constrained by thermodynamics, and if so to what extent? 
How fast can neural computation be to maintain thermal balance and
physiological range of intracellular ionic concentrations?

This article answers these questions by finding the power generated
by sodium metabolic pumps and relating this power to the thermal
and neuroanatomical properties of brain tissue. From this, we
determine theoretical thermal bounds on fiber dimensions, and compare 
them with empirical data. These bounds enable us to estimate the
upper theoretical limits on the density of brain wiring.

\vspace{2.5cm}

\section{Methods}

\subsection{Voltage and Na$^{+}$-K$^{+}$ dynamics.}

Na$^{+}$ and K$^{+}$ are two major ions affecting neural membrane
dynamics (Kandel et al, 1991; Koch, 1998) and its energetics
(Astrup et al, 1981; Erecinska and Silver, 1989; Rolfe and Brown, 1997;
Ames, 2000). The energy consuming Na$^{+}$/K$^{+}$ pump affects membrane 
electrical properties, because it pumps out 3 Na$^{+}$ and pumps in 2 
K$^{+}$ per 1 ATP molecule consumed (Kandel et al, 1991).
The equations governing membrane and Na$^{+}$/K$^{+}$-ATP activities
of a neuron are given by:

\begin{eqnarray}
CS \frac{dV}{dt} = - g_{Na}S(V-V_{Na}) - g_{K}S(V-V_{K})
- g_{L}S(V-V_{L}) - I_{p} - I_{s}, \nonumber \\
U_{n}F \frac{d[\mbox{Na}]}{dt} = - g_{Na}S(V-V_{Na}) - 3I_{p}
 - \alpha I_{s}, \nonumber \\
U_{n}F \frac{d[\mbox{K}]}{dt} = - g_{K}S(V-V_{K}) + 2I_{p}
 - \beta I_{s}, \nonumber \\
\end{eqnarray}\\
where $V$ is the membrane potential, $C$ is the membrane capacitance
per unit area, $S$ is neuron's membrane surface area (primary axons
and dendrites), $V_{Na}$ and $V_{K}$ are Na$^{+}$ and K$^{+}$ 
reversal potentials, $g_{Na}$ and $g_{K}$ are voltage-dependent 
Na$^{+}$ and K$^{+}$ conductances per unit membrane area, $g_{L}$ 
is the leak conductance and $V_{L}$ is the reversal potential 
corresponding to the leak current. The parameter $U_{n}$ denotes 
neuron's volume, $F$ is the Faraday constant. The current
$I_{p}$ is the Na$^{+}$/K$^{+}$ pump current given by 
$I_{p}= AS [\mbox{Na}]^{k}/([\mbox{Na}]^{k}+\theta^{k})$ where $\theta= 20.0$
mM, $A$ is the maximal pump current per membrane surface area, and
$k$ is the Hill constant. The symbol [...] denotes 
intracellular ionic concentration, $I_{s}$ is the synaptic current, 
and $\alpha$, $\beta$ are voltage dependent proportionality parameters 
determined below.

Voltage dependent conductances are represented by 
$g_{Na}= m^{3}h\overline{g}_{Na}$ and $g_{K}= n^{4}\overline{g}_{K}$,
where $\overline{g}_{Na}$ and $\overline{g}_{K}$ are the maximal Na$^{+}$
and K$^{+}$ conductances. The gating variables $h$ and $n$ obey
the standard kinetic equation $dx/dt= \alpha_{x}(1-x)-\beta_{x}x$,
(for $x=h, n$), and the fast variable $m$ is set to its equilibrium
value $\alpha_{m}/(\alpha_{m}+\beta_{m})$. Voltage dependences of
the parameters $\alpha_{x}$ and $\beta_{x}$ were chosen, with a slight 
modification, as in the Traub-Miles model (1991), which describes 
a pyramidal neuron in the hippocampus. Specifically, for sodium 
channels: $\alpha_{m}= 0.32(V+54)/\left(1-\exp(-(V+54)/4)\right)$,
$\beta_{m}= 0.28(V+27)/\left(\exp((V+27)/5)-1\right)$,
$\alpha_{h}= 0.128\exp(-(V+50)/18)$,
$\beta_{h}= 4.0/(1+\exp(-(V+27)/5)$, and for potassium channels
$\alpha_{n}= 0.044(V+52)/\left(1-\exp(-(V+52)/5)\right)$,
$\beta_{n}= 0.5\exp(-(V+57)/40)$, where $V$ is expressed in mV.

It is assumed that the neuron represented by Eq. (1) is embedded in
the network of neurons firing with an average firing rate $f$.
The synaptic current of a single synapse is 
$q g_{s}e^{-t/\tau_{s}}(V-V_{s})$, where $q$ is the release
probability, $g_{s}$ is the maximal synaptic conductance
(it is assumed that the rising phase of synaptic conductance is much
faster that its decaying phase that is characterized by the time
constant $\tau_{s}$), and $V_{s}$ is the reversal potential for
synapses. Since the majority of synapses in the mammalian brain are 
excitatory and glutamate (Braitenberg and Schuz, 1998), we have 
$V_{s}= 0$ (Koch, 1998). The average synaptic conductance is 
$f\int_{0}^{1/f} dt\; q g_{s}e^{-t/\tau_{s}}$, which in
the physiologically valid limit $f\tau_{s} \ll 1$ yields 
$q g_{s}f\tau_{s}$. Neurophysiological data indicate 
(Markram et al, 1997) that the release probability $q$ is frequency 
dependent. Here, we follow Dayan and Abbott (2001) and assume that 
synaptic depression modifies the release probability according to
$q(f)= q_{0}/(1 + \gamma\tau_{d}f)$, where $q_{0}$ is the release
probability at 0 Hz, $\tau_{d}$ is the depression time constant, and 
$\gamma$ is the parameter controlling the degree of depression
(for $\gamma= 0$ lack of depression, for $\gamma= 1$ maximal depression).
Thus, the total average synaptic current is 
$I_{s}= fq_{0}Mg_{s}\tau_{s}(V-V_{s})/(1+\gamma\tau_{d}f)$,
where $M$ is the number of synapses (or presynaptic neurons) per neuron.
Fluctuations around the average $I_{s}$ cause the neuron to fire
an action potential with an average firing rate $f$.
Glutamate synapses with non-NMDA receptors are used more frequently
for regular transmission (NMDA receptors are blocked by Mg$^{2+}$; 
Kandel et al, 1991) and therefore only these are considered here. Synaptic 
current of non-NMDA type is composed almost exclusively of Na$^{+}$ 
and K$^{+}$ ions, which is the reason for the presence of $\alpha I_{s}$
and $\beta I_{s}$ terms in the dynamics of [Na] and [K]
in Eq. (1), where $\alpha + \beta = 1$.

The parameter $\alpha$ can be determined in a few steps. 
First, we can write the synaptic current $I_{s}$ as
$I_{s}= \tilde{g}_{Na}(V-V_{Na}) + \tilde{g}_{K}(V-V_{K})$,
where $\tilde{g}_{K}$ and $\tilde{g}_{Na}$ are K$^{+}$ and Na$^{+}$ 
conductances (voltage independent) at glutamatergic synapses. 
Since for $V= V_{s}= 0$ the synaptic current $I_{s}= 0$, we have
$\tilde{g}_{K}= - \tilde{g}_{Na} V_{Na}/V_{K}$.
Second, from the condition $\alpha I_{s}= \tilde{g}_{Na}(V-V_{Na})$, 
we obtain $\alpha= \tilde{g}_{Na}(V-V_{Na})/(\tilde{g}_{Na}(V-V_{Na})
+\tilde{g}_{K}(V-V_{K}))$, which leads to 
$\alpha= V_{K}(V-V_{Na})/\{V(V_{K}-V_{Na})\}$. Note that $\alpha$
is voltage dependent. Consequently, the synaptic contribution to changes 
in intracellular Na$^{+}$ concentration is given by 
$\alpha I_{s}= q(f)fMg_{s}\tau_{s}(V-V_{Na})V_{K}/(V_{K}-V_{Na})$.
The synaptic contribution to changes in intracellular K$^{+}$ is
$\beta I_{s}= -q(f)fMg_{s}\tau_{s}(V-V_{K})V_{Na}/(V_{K}-V_{Na})$.

Equation for voltage dynamics in Eq. (1) assumes that the membrane is
equipotential, i.e., axons and dendrites have on average equal potentials.
This assumption neglects spatio-temporal effects associated with action 
potential propagation. The validity of this approximation is discussed
in the Discussion section and in Appendix A.

Adding appropriately sides in Eq. (1) we obtain the ``charge conservation''
equation:

\begin{equation}
U_{n}F\frac{d}{dt}\left( [Na] + [K] \right)=
CS\frac{dV}{dt} + g_{L}(V-V_{L}).
\end{equation}\\
This equation reflects the fact that non-zero membrane potential is caused 
by concentration gradients of charged ions across membrane, and temporal
changes in potential are directly related to the temporal changes in ionic
concentrations. This temporal relationship during an initial phase of
an action potential is discussed in the Results section. Here, we focus
instead on the long-term relation between intracellular [Na] and [K]. 
In this regime and for firing rates $ < 100$ Hz (duration of action
potential and voltage recovery $\sim 10-15$ msec), the average of $V$ is 
close to its resting potential $V_{o}$, which in turn is very close to 
$V_{L}$. Thus, on the long-time scale, the right hand side of Eq. (2) 
is close to 0, which implies that [Na] + [K] $\approx const$. That is,
changes in [Na] directly determine changes in [K], and practically there 
is no need to solve the equation for the potassium dynamics.
Values of neurophysiological parameters used in this paper are presented 
in Table 1.

\subsection{Neuroanatomical relationships.}

In studying metabolic and thermodynamic properties of brain tissue
the following neuroanatomical relations were used: 
(i) volume density of synapses $\rho_{s}= NM/U_{g}$
is brain size independent, where $N$ is the number of neurons and
$U_{g}$ is the cortical gray matter volume
(Braitenberg and Schuz, 1998; DeFelipe et al, 2002); (ii) the fraction
$1-\phi$ of the gray matter volume taken by fibers (axons and dendrites)
is approximately constant and about 2/3 (Braitenberg and Schuz, 1998); 
(iii) white matter volume $U_{w}$ scales with
gray matter volume $U_{g}$ as $U_{w}= 0.166 U_{g}^{1.23}$ (Zhang and
Sejnowski, 2000), where $U_{g}$ and $U_{w}$ are expressed in cm$^{3}$.

The relationship between neuron's surface area $S$ and gray matter volume
$U_{g}$ can be determined as follows. The gray matter volume is
$U_{g}= NU_{n} + \phi U_{g}$, where neuron's volume 
$U_{n}= (l_{a}d_{a}^{2} + l_{d}d_{d}^{2})\pi/4$ with $l_{a}$, $l_{d}$
denoting average (unmyelinated) axon and dendrite length in the gray 
matter and $d_{a}$, $d_{d}$ denoting their corresponding diameters 
(neuron's soma is
neglected as it is small). The parameter $\phi$ is the fraction of volume
taken by non-fibers (synapses, blood vessels, etc). The neuron's surface 
area is $S= \pi(l_{a}d_{a}+l_{d}d_{d})$. We define ``an effective'' fiber
diameter $d$ as

\begin{equation}
d\equiv (l_{a}d_{a}^{2} + l_{d}d_{d}^{2})/
(l_{a}d_{a}+l_{d}d_{d}),
\end{equation}\\
which is proportional to the neuron's volume to surface ratio, i.e.,
$d= 4U_{n}/S$. In fact, if volumes of axons and dendrites are equal, 
which seems to be neuroanatomically valid in the gray matter 
(Braitenberg and Sch{\"u}z, 1998; Chklovskii et al, 2002), 
then $d$ represents a harmonic mean of axonal $d_{a}$ and dendritic 
$d_{d}$ diameters, that is $d= 2d_{a}d_{d}/(d_{a}+d_{d})$. This formula
can be derived by noting that axon length $l_{a}= l_{d}(d_{d}/d_{a})^{2}$
(equal volumes of axons and dendrites), and substituting $l_{a}$ to
Eq. (3). Note that for constant $d_{d}$, the effective diameter $d$ 
decreases with decreasing $d_{a}$. In the limit $d_{a}/d_{d} \ll 1$, we
have $d\approx 2d_{a}$.

For neuroanatomical values of $d_{a}=0.3$ $\mu$m and $d_{d}=0.9$ $\mu$m
in mouse (Braitenberg and Sch{\"u}z, 1998), we obtain the effective fiber
diameter $d=0.45$ $\mu$m. Since the average values of cortical axon and 
dendrite diameters are roughly invariant with respect to brain size 
(i.e., they should be more or less the same for mouse and human; 
Braitenberg and Sch{\"u}z, 1998), we can
expect that the effective fiber diameter $d$ should not change with
brain size either. Therefore, we will use the above value of $d= 0.45$ 
$\mu$m for illustrative purposes throughout the paper. Apart from that,
occasionally we will also show results for the smallest experimentally 
known axons of diameter 0.1 $\mu$m corresponding to $d= 0.18$ $\mu$m 
(for $d_{d}= 0.9$ $\mu$m), and for the smallest molecularly feasible 
axons (neurites) of diameter 0.06 $\mu$m (axons without Natrium channels 
that lack action potentials; Faisal et al, 2005), corresponding to
$d= 0.11$ $\mu$m (also for $d_{d}= 0.9$ $\mu$m).

With the definition of $d$ we can relate the total surface
area of neurons, $NS$, to the gray matter volume $U_{g}$ as

\begin{equation}
NS= 4(1-\phi)U_{g}/d, 
\end{equation}\\
and relate the surface density of synapses along axons and dendrites,
$M/S$, to synaptic density $\rho_{s}$ as

\begin{equation}
\frac{M}{S}= \frac{\rho_{s}d}{4(1-\phi)}.
\end{equation}\\
These relations are used below to rewrite Eq. (1) in a more convenient
form for numerical and analytical calculations.

\subsection{Explicit voltage and ionic dynamics.}

With the modifications and interdependencies between parameters
described in Subsections 2.1 and 2.2 we can rewrite Eq. (1) as:

\begin{eqnarray}
C\frac{dV}{dt}= - g_{Na}(V-V_{Na}) - g_{K}(V-V_{K}) - g_{L}(V-V_{L})
\nonumber \\
   - \frac{A[\mbox{Na}]^{k}}{[\mbox{Na}]^{k}+\theta^{k}}
    - \frac{q(f)f\rho_{s}dg_{s}\tau_{s}V}{4(1-\phi)}
 + synaptic \; fluct.
\nonumber \\
\frac{dF}{4}\frac{d[\mbox{Na}]}{dt}= - \left(g_{Na}
  + \frac{q(f)f\rho_{s}dg_{s}\tau_{s}V_{K}}
  {4(1-\phi)(V_{K}-V_{Na})}\right)(V-V_{Na})
   - \frac{3A[\mbox{Na}]^{k}}{[\mbox{Na}]^{k}+\theta^{k}}
\nonumber \\
\frac{dF}{4}\frac{d[\mbox{K}]}{dt}=  - \left(g_{K}
  - \frac{q(f)f\rho_{s}dg_{s}\tau_{s}V_{Na}}
  {4(1-\phi)(V_{K}-V_{Na})}\right)(V-V_{K})
   + \frac{2A[\mbox{Na}]^{k}}{[\mbox{Na}]^{k}+\theta^{k}},
\end{eqnarray}\\
where the release probability modulated by synaptic depression is 
$q(f)= q_{0}/(1 + \gamma\tau_{d}f)$ (Dayan and Abbott, 2001).
These equations are solved both numerically (using the 4th-order
Runge-Kutta method) and approximately analytically. Fast
synaptic fluctuations denoted symbolically in the top equation
of Eq. (6) cause the neuron to fire stochastically with the
average frequency $f$ (i.e. on average, every $1/f$ seconds the
voltage is set to $-20$ mV).

\subsection{Electric power generated by Na$^{+}$/K$^{+}$-ATP pumps.}

Since the metabolic rate in white matter is 3-4 times lower than
in gray matter (Siesjo, 1978; Karbowski, 2007), the white matter 
contribution to the total brain metabolic power is neglected.
The average power $P_{ATP}$ dissipated by $N$ neurons in the gray matter 
is given by

\begin{eqnarray}
P_{ATP}= \frac{N}{\Delta t} \int_{0}^{\Delta t} dt \; \left( -3I_{p}(V-V_{Na}) 
+ 2I_{p}(V-V_{K})\right), 
\end{eqnarray}\\
where the integral represents the electrical work performed by
Na$^{+}$/K$^{+}$-ATP pumps during the long time $\Delta t$, much larger than 
the average interspike interval $1/f$. This work goes for removing 3 Na$^{+}$ 
ions and importing 2 K$^{+}$ ions against their electrochemical gradients, 
which cost 1 ATP molecule (Kandel et al, 1991). Opposite signs in front of 
$3I_{p}$ and $2I_{p}$ indicate the fact that Na$^{+}$ and K$^{+}$ ions 
move in opposite directions through the membrane.

Integral in Eq. (7) can be estimated by noting that in the long-time 
limit the pump current $I_{p}$ assumes its average value (with some 
fluctuations around, see the Results) given by 
$I_{p,av}= AS [\mbox{Na}]^{k}_{av}/([\mbox{Na}]^{k}_{av} + \theta^{k})$, 
where $[\mbox{Na}]_{av}$ is the average long-term
sodium concentration (determined in the Results section). In this
limit, the average value of voltage $V(t)$ is approximately equal to
its resting value $V_{o}$. Thus, the electric power $P_{ATP}$ is

\begin{equation}
P_{ATP}\approx \frac{NSA[\mbox{Na}]_{av}^{k}}
{([\mbox{Na}]^{k}_{av} + \theta^{k})}
\left(3V_{Na}-2V_{K}-V_{o}\right), 
\end{equation}\\
where $V_{Na}= (RT/F)\ln([\mbox{Na}]_{ex}/[\mbox{Na}]_{av})$ and
$V_{K}= (RT/F)\ln([\mbox{K}]_{ex}/[\mbox{K}]_{av})$, with the extracellular 
Na$^{+}$ and K$^{+}$ concentrations (Hille, 2001): 
$[\mbox{Na}]_{ex}= 145$ mM, and $[\mbox{K}]_{ex}= 4$ mM.

\subsection{Mechanisms of brain cooling.}

The major contribution to the cerebral metabolic rate or heat constitute
Na$^{+}$/K$^{+}$-ATPase (Astrup et al, 1981; Erecinska and Silver, 1989;
Rolfe and Brown, 1997; Ames, 2000). Therefore, in this paper, the heat
coming from other reactions such as glycolysis (conversion of glucose
to ATP) is assumed to be less important and hence it is neglected in the
estimates. In deep brain regions heat generated by hydrolysis of ATP 
molecules and coupled to it activity of Na$^{+}$/K$^{+}$
pumps is transfered by conduction and circulating cerebral blood flow,
whereas at the surface (scalp) the heat is removed mostly by 
convection/conduction and radiation (Fig. 1). 
Heat conduction (convection) is associated with the 
existence of a temperature gradient between the brain or scalp and the 
external environment. Heat radiation is a quantum mechanical property of 
every physical object having temperature above absolute zero. Transfer of 
heat to the circulating cerebral blood in deep brain regions is possible 
because incoming blood (artery) has a slightly lower temperature than 
the brain tissue in this region (Hayward and Baker, 1968; 
Nybo et al, 2002; Kiyatkin, 2007). In general, the incoming 
blood temperature corresponds to the core body temperature, which is
assumed to be a constant because it changes on time scales that are 
much slower than changes in brain temperature (Kiyatkin, 2007).

The equation governing heat balance and spatial distribution of brain
temperature $T(r)$ has the form (Nelson and Nunneley, 1998;
van Leeuwen et al, 2000; Sukstanskii and Yablonskiy, 2006):

\begin{eqnarray}
\rho_{br}c_{br}\frac{\partial T}{\partial t}= \kappa\frac{\partial^{2} T}
{\partial r^{2}} - \rho_{bl}c_{bl}\mbox{CBF}(T-T_{bl}) + G_{ATP}/U_{br},
\end{eqnarray}\\
where $\kappa$ is the arithmetic mean of the thermal conductance of 
brain tissue (including cerebrospinal fluid, skull, and scalp at the edge),
$\rho_{bl}$ and $c_{bl}$ denote blood's density and specific heat, CBF
is the cerebral blood flow rate expressed in sec$^{-1}$, $T_{bl}$ is
the incoming blood (arterial) temperature equivalent to body core
temperature, and $G_{ATP}$ is the heat (per time unit) generated in the 
gray matter due to Na$^{+}$/K$^{+}$-ATPase. In general, $G_{ATP} > P_{ATP}$
because $P_{ATP}$ is the useful work (per time unit) performed due to
hydrolysis of ATP and release of $G_{ATP}$ of free energy (per time
unit). In Sec. 3.5 we estimate the relative magnitude of $G_{ATP}$ and
$P_{ATP}$, i.e., we calculate the efficiency of the sodium pump.
The parameter $U_{br}$ denotes brain volume, i.e., $U_{br}= U_{g} + U_{w}$
(brain geometry is modelled as half of a ball; Fig. 1). The parameters
$\rho_{br}$ and $c_{br}$ denote brain tissue density and specific heat
and they are approximately equal to $\rho_{bl}$ and $c_{bl}$, respectively.
Only steady-state regime of Eq. (9), i.e. $\partial T/\partial t= 0$,
is considered below.

The boundary condition imposed on Eq. (9) has the following form:

\begin{eqnarray}
\kappa \frac{\partial T}{\partial r}|_{r=R} = 
- \left(\sigma_{SB}(T_{sc}^{4}-T_{o}^{4}) + \eta(T_{sc}-T_{o})\right),
\end{eqnarray}\\
where $R$ is the brain's radius given by $R= (3U_{br}/2\pi)^{1/3}$,
$\sigma_{SB}$ is the Stefan-Boltzmann constant, $\eta$ is the heat
convection/conduction coefficient between the scalp and the outside 
environment, and $T_{sc}$ and $T_{o}$ denote the scalp and environment
temperatures. The left hand side of Eq. (10) corresponds to the heat rate 
removed from the scalp per surface area. The right hand side of Eq. (10)
is the sum of scalp radiation (term $\sim T_{sc}^{4}$) and scalp 
convection/conduction (term $\sim T_{sc}$). The presence of the term
proportional to $T_{o}^{4}$ is a consequence of the fact that scalp
not only radiates energy to the environment, but it also receives some
radiation from it.

The heat removed from the brain through the conduction is given by

$\dot{Q}_{c}= -\kappa \int d^{3}r \; \partial^{2} T/\partial r^{2}
= -2\pi R^{2}\kappa \partial T/\partial r |_{r=R}$. The heat removed
by the circulating cerebral blood is given by $\dot{Q}_{bl}=
\rho_{bl}c_{bl}\mbox{CBF} \int d^{3}r \; (T(r)-T_{bl})$. 
Therefore, by integrating Eq. (9) for the whole brain volume, 
at the steady-state, we obtain:

\begin{eqnarray}
G_{ATP}= \dot{Q}_{c} + \dot{Q}_{bl},
\end{eqnarray}\\
where the conduction term $\dot{Q}_{c}$ is a sum of scalp 
convection/conduction and scalp radiation, i.e.,
$\dot{Q}_{c}= \dot{Q}_{cv} + \dot{Q}_{r}$, with 
$\dot{Q}_{cv}= 2\pi R^{2}\eta(T_{sc}-T_{o})$ and
$\dot{Q}_{r}= 2\pi R^{2}\sigma_{SB}(T_{sc}^{4}-T_{o}^{4})$.
Values of all thermodynamic parameters are presented in Table 1.

\vspace{2.5cm}


\section{Results}

\subsection{Sodium influx during a single action potential.}

When synaptic fluctuations cause the neuron to fire an action potential,
the Na$^{+}$ concentration first rise and then slowly decays to its 
equilibrium value, due to the workings of the Na$^{+}$/K$^{+}$ pump. 
Sodium influx during a single action potential can be estimated using 
Eqs. (2) and (6). It is composed of the two contributions (Fig. 2A): 
Na$^{+}$ influx during a rising phase of sodium conductance $g_{Na}$ and 
voltage $V$, and Na$^{+}$ influx during a decline phase of $g_{Na}$ and 
$V$. The total influx is given by (see Appendix A):

\begin{eqnarray}
\Delta[\mbox{Na}]_{o}\approx \frac{4(C + \delta C)}{Fd}(V_{Na}-V_{o}),
\end{eqnarray}\\
where $\delta C$ denotes ``correction'' to the effective capacitance 
coming from the prolonged Na$^{+}$ channels activation and is given by 
$\delta C\approx 0.064\overline{g}_{Na}\tau_{o}(V_{Na}-0.6V_{K})/
(V_{Na}-V_{o})$. Value of $\tau_{o}$ is taken from numerical simulations 
of Eq. (6), and is approximately $\tau_{o}\approx 0.4$ msec. 
This number enables us to determine the parameter $\delta C$ in Eq. (12), 
which we find to be $\delta C\approx 2.4$ $\mu$F/cm$^{2}$ for typical 
resting values of voltages: 
$V_{Na}= 68$ mV, $V_{K}= -100$ mV 
(for [Na]= 12 mM, [K]= 155 mM; Hille, 2001) and $V_{o}= -67$ mV. 
It should be remembered, however, that $\delta C$ is not a constant, but 
it varies slightly depending on the level of intracellular sodium 
concentration [Na] via $V_{Na}$ and $V_{K}$ (see Fig. 2B).

The immediate conclusion from Eq. (12) is that the amplitude of the 
sodium influx increases with decreasing the effective fiber diameter $d$,
and this agrees with a direct numerical integration of Eq. (6),
as is shown in Fig. 2C. Based on the above values, we find from Eq. (12) 
that for $d= 0.45$ $\mu$m (harmonic mean of average axon and
dendrites diameters in mouse; Sec. 2.2) the sodium influx 
$\Delta[\mbox{Na}]_{o}\approx 0.42$ mM, which corresponds to 
$\sim 2.5\cdot 10^{5}$ Na$^{+}$ ions per $\mu$m$^{3}$.
For a mouse neuron with equal volume of axons and dendrites, and with 
$l_{a}= 4$ cm and $d_{a}= 0.3$ $\mu$m, this gives the total
influx of $14.2\cdot 10^{8}$ Na$^{+}$ ions. Direct numerical integration 
of Eq. (6) for $d=0.45$ $\mu$m yields a similar sodium influx, 
$\Delta[\mbox{Na}]_{o}\approx 0.47$ mM, which corresponds to the total 
influx of $15.9\cdot 10^{8}$ Na$^{+}$ ions for a mouse neuron.

The influx of $14.2-15.9\cdot 10^{8}$ Na$^{+}$ ions during an isolated 
action potential obtained above is comparable to the estimate of Attwell 
and Laughlin (2001), who used a different, phenomenological, approach and 
obtained a slightly lower value of $11.5\cdot 10^{8}$ Na$^{+}$ ions. The 
difference can be attributed to the differences in the assumption regarding 
spatial properties of the membrane potential and the amplitude of 
depolarization during an action potential. In this paper, it is assumed 
that the membrane is equipotential, i.e., axons and dendrites are 
depolarized by the same amount (135 mV), whereas Attwell and Laughlin 
(2001) assume that dendrites 
are 50 $\%$ less polarized than axons (dendrites 50 mV, axons 100 mV).
Below, it is shown that the formula for Na$^{+}$ charge influx used 
by these authors is equivalent to the formula (12), 
if we assume that the axon depolarization $\Delta V_{a}$ during
an action potential is the same as the dendrite depolarization
$\Delta V_{d}$. The total charge influx $Q_{AL}$ in Attwell and 
Laughlin (2001) is 
$Q_{AL}= 4\pi C(l_{a}d_{a}\Delta V_{a} + l_{d}d_{d}\Delta V_{d})$,
if we neglect a small soma contribution. The prefactor of 4 was chosen by 
these authors to a large extent arbitrary, and it comes from 
the effect of simultaneous activation of Na$^{+}$ and K$^{+}$ channels,
which is analogous to the presence of the $\delta C$ contribution in Eq. (12). 
Now, assuming that $\Delta V_{d}= \Delta V_{a}\equiv \Delta V$, and noting 
that $Q_{AL}= FU_{n}\Delta[\mbox{Na}]_{AL}$, where $\Delta[\mbox{Na}]_{AL}$ 
is the sodium influx in the Attwell and Laughlin (2001) formulation, 
we obtain $\Delta[\mbox{Na}]_{AL}= 16C\Delta V/(Fd)$, where $d$ is the 
effective fiber diameter defined in Eq. (3). 
The formula for $\Delta[\mbox{Na}]_{AL}$ is very similar 
to the formula (12), except for the numerical factor in front, which is 
13.6 in Eq. (12). This results from the fact that numerically 
$\delta C= 2.4C$ for low firing rates or intracellular sodium 
concentrations (Fig. 2B).
It is also interesting to note that the Attwell-Laughlin (2001) formula
does not account for changes in the intracellular Na$^{+}$ concentration
due to repetitive firing, as opposed to Eq. (12) that includes such 
changes through $\delta C$. In this sense Eq. (12) extends the 
phenomenological approach of these authors into the broader range 
of frequencies.

\subsection{Sodium build-up due to repetitive firing.}

When the neuron fires repeatedly, its intracellular sodium accumulates
with every spike because Na$^{+}$/K$^{+}$ pump is slow and cannot
remove all Na$^{+}$ promptly (Fig. 3A,B). Since every spike introduces
$U_{n}F\Delta[\mbox{Na}]_{o}$ of sodium electric charge, and this process
is fast, we can approximate Eq. (6) for the Na$^{+}$ dynamics as

\begin{eqnarray}
U_{n}F\frac{d[\mbox{Na}]}{dt}\approx I_{o} 
 + U_{n}F\Delta[\mbox{Na}]_{o}\sum_{i} \delta(t-t_{i})
   - \frac{3AS[\mbox{Na}]^{k}}{[\mbox{Na}]^{k}+\theta^{k}},
\end{eqnarray}\\
where $I_{o}$ is the sodium current associated with sodium channels 
and synaptic contribution at rest, and it is given by
$I_{o}= \left\{ g_{Na,o} + q(f)f\rho_{s}dg_{s}\tau_{s}V_{K}/(4(1-\phi)
(V_{K}-V_{Na}))\right\}S(V_{Na}-V_{o})$. The delta functions present
in Eq. (13) represent spikes of Na$^{+}$ influx at times ${t_{i}}$.
Simulation of Eq. (13) is shown in Fig. 3C, and it resembles the
simulation of the original Eq. (6) (see Fig. 3A).

We can further simplify Eq. (13) at the long-time limit, in which
we substitute for the right hand side of Eq. (13) its temporal
average. In particular, the temporal average of $\sum_{i} \delta(t-t_{i})$
is equal to the firing rate $f$, and thus

\begin{eqnarray}
U_{n}F\frac{d\overline{[\mbox{Na}]}}{dt}\approx 
I_{o} + fU_{n}F\Delta[\mbox{Na}]_{o}
- \frac{3AS\overline{[\mbox{Na}]}^{k}}{\overline{[\mbox{Na}]}^{k}+\theta^{k}},
\end{eqnarray}\\
where $\overline{[\mbox{Na}]}$ is the temporal average of [Na]. A comparison
of the time dependence of $\overline{[\mbox{Na}]}$ from Eq. (14) with the 
time dependence of [Na] coming from Eq. (13) is presented in Fig. 3C. 
The equilibrium value of $\overline{[\mbox{Na}]}$, denoted as [Na]$_{av}$ 
is determined from the condition $d\overline{[\mbox{Na}]}/dt= 0$, with 
the help of Eq. (12). [Na]$_{av}$ satisfies the following equation:

\begin{eqnarray}
 \frac{3A[\mbox{Na}]_{av}^{k}}{[\mbox{Na}]_{av}^{k}+\theta^{k}}
\approx I_{o}/S + f(C+\delta C)(V_{Na}-V_{o}).
\end{eqnarray}\\
Note that the right hand side of Eq. (15) also depends on [Na]$_{av}$
through $V_{Na}$ and $\delta C$. Thus, we can find [Na]$_{av}$ only
numerically, either from Eq. (15) or from a direct simulation of 
Eq. (14).

In Fig. 4, we compare the dependence of [Na]$_{av}$ on firing rate that
comes from Eq. (15) with the dependence that comes from a direct numerical
integration of Eq. (6). Overall, the formula 
(15) provides a relatively good approximation to the numerical solution,
especially for low firing rates (Fig. 4). For this reason, it is used 
in the following sections to approximate metabolic expenditure and 
thermal changes in the brain tissue.

It is interesting to note that the average [Na]$_{av}$ is weakly
dependent on fiber diameter, especially for small firing rates 
(Fig. 4). The reason for this is that only the current $I_{o}$ in Eq. (15)
contains synaptic contribution with $d$, and for low frequencies
$I_{o}$ is small.

It should be kept in mind, however, that sodium fluctuations around 
[Na]$_{av}$ do grow with decrease in fiber diameter (Fig. 3B), because
for thin fibers the amplitude of sodium influx is large 
($\Delta[\mbox{Na}]_{o} \sim 1/d$) and the relaxation time constant 
is short ($\tau \sim d$; see Appendix B).
Enhanced intracellular sodium fluctuations can effectively reduce the 
concentration gradients across neuron's membrane to values close to zero,
and this would have a devastating effect on neuron's functionality if 
high firing rates were maintained for a prolonged period of time. 
For example, for $d=0.11$ $\mu$m corresponding to the smallest physically 
possible axons of $d_{a}= 0.06$ $\mu$m (Faisal et al, 2005) the 
intracellular sodium concentration can occasionally peak to extracellular 
levels (145 mM; Hille 2001) just for 40 Hz of repetitive firing. 
For $d= 0.45$ $\mu$m (corresponding to $d_{a}= 0.3$ $\mu$m) this takes
place for $\sim 50$ Hz. These considerations suggest that very thin 
fibers are not beneficial for neuron's electrical properties.

\subsection{ATP utilization rate of Na$^{+}$/K$^{+}$ pump, glucose
metabolism, and firing rate in mammals.}

For pumping out Na$^{+}$ and pumping in K$^{+}$, the Na$^{+}$/K$^{+}$
pump uses ATP molecules. The average ATP utilization rate of a single
neuron per its surface area $S$ is equal to $I_{p,av}/(FS)$. 
Generally, it increases with 
[Na]$_{av}$ and thus with the firing rate $f$ in a non-linear fashion 
(Fig. 5A). For low frequencies $f$ there exist a linear regime, while 
for high $f$, the ATP rate saturates reaching its maximum value 
$A/F$ (Fig. 5A). Typical values of ATP utilization for
frequencies in the range $1-10$ Hz are $(0.6-9)\cdot 10^{4}$ ATP
molecules per $\mu$m$^{2}$ per second. In the linear regime, the 
increase of firing rate by 1 Hz leads to the increase of the ATP rate
by about $1.7\cdot 10^{-4}$ $\mu$mol/(cm$^{2}$sec). It is important to 
point out that in previous phenomenological models of
ATP utilization rate (Attwell and Laughlin 2001; Lennie, 2003) only
linear regime was assumed in estimations. Thus, the present more
detailed model extends these calculations into the non-linear regime.

The average firing rate in mammalian brains can be determined indirectly
from the cerebral glucose utilization rates CMR$_{glu}$ (expressed in
mol/(cm$^{3}\cdot$s) ). If we assume that the ATP activity of the neural 
pumps constitutes the major contribution to the gray matter metabolism 
(Astrup et al, 1981; Erecinska and Silver, 1989;
Rolfe and Brown, 1997; Ames 2000), then we can relate directly the ATP
rate or the pump current to CMR$_{glu}$. Since 31 ATP molecules are
produced per 1 glucose molecule (Rolfe and Brown, 1997), we have 
for the whole gray matter the following equality 
$31 U_{g} \mbox{CMR}_{glu}= N I_{p,av}/F$, from which we obtain, using 
Eq. (4), that

\begin{eqnarray}
\mbox{CMR}_{glu}= \frac{4(1-\phi)A[\mbox{Na}]^{k}_{av}}
{31Fd([\mbox{Na}]^{k}_{av}+\theta^{k})}.
\end{eqnarray}\\
We can write CMR$_{glu}$ in an equivalent form, which contains 
neurophysiological parameters explicitly. Using Eq. (15), we obtain:

\begin{eqnarray}
\mbox{CMR}_{glu}\approx \frac{4(1-\phi)}{93Fd}\left( I_{o}/S + f(C+\delta C)
(V_{Na}-V_{o})\right).
\end{eqnarray}\\
Dependence of CMR$_{glu}$ on firing rate $f$ is plotted in Fig. 5B,
and it is practically the same as the dependence of $I_{p,av}$ on
$f$, as the two quantities CMR$_{glu}$ and $I_{p,av}$ are proportional.
The non-linear part of the dependence comes from the fact that $\delta C$
and $V_{Na}$ decrease for high frequency (via [Na]$_{av}$). The
CMR$_{glu}$ vs. $f$ relationship (Eq. 17) enables us to find average 
frequencies for several mammalian species for which empirical values of 
CMR$_{glu}$ are known (Fig. 5C). In general, the average firing rates are 
rather low, from $\sim$ 1.7 Hz for human to $\sim$ 6.2 Hz for mouse 
(Table 2). Moreover, estimated in such a way average firing rates scale 
with gray matter volume with an exponent of $-0.15$ (Fig. 5C), implying 
that average activity in larger brains is slower than in smaller brains.

\subsection{Biphasic dependence of pump power on frequency.}

The electric power generated by the Na$^{+}$/K$^{+}$ pump in a single neuron
(per surface area) is determined in two ways. First, from a direct numerical 
integration of Eq. (7) with time dependent voltage $V$ and pump current 
$I_{p}$. Second, from the approximate analytical formula (8) with the help 
of derived Eq. (15). Both methods yield similar results (Fig. 6A), 
which indicates that the approximation (8) is reliable, especially for
low firing rates. We can rewrite Eq. (8) for the total electric power 
generated in the gray matter in a more convenient form using Eq. (4). 
The result is

\begin{equation}
P_{ATP}\approx \frac{4(1-\phi)U_{g}A[\mbox{Na}]_{av}^{k}}
{d([\mbox{Na}]^{k}_{av} + \theta^{k})}
\left(3V_{Na}-2V_{K}-V_{o}\right).
\end{equation}\\
Alternatively, we can use Eq. (15) to relate the pump current to the
neurophysiological parameters. In this way, we obtain:

\begin{eqnarray}
P_{ATP}\approx \frac{4(1-\phi)U_{g}}{3d}
\left(3V_{Na}-2V_{K}-V_{o}\right)(V_{Na}-V_{o})
\nonumber \\
\times \left( g_{Na,o} + \frac{fq\rho_{s}dg_{s}\tau_{s}V_{K}}
{4(1-\phi)(V_{K}-V_{Na})} + f(C+\delta C) \right).
\end{eqnarray}\\
Note that synaptic depression via $q$ reduces the power $P_{ATP}$.
In Eq. (19) the first term in the large bracket represents a very 
small sodium influx at rest, the second term corresponds to the background
dendritic synaptic activity, and the last term comes from Na$^{+}$ influx
due to action potentials. The relative contribution of these 3 elements
to $P_{ATP}$ depends on firing rate $f$. For example, for $f=1.7$ Hz,
corresponding to the estimate for human brain (Table 2), sodium influx
at rest constitutes 5 $\%$, background dendritic synaptic activity
15 $\%$, and Na$^{+}$ influx due to action potentials yields 80 $\%$
of the total power.

Fig. 6A indicates that $P_{ATP}$ depends biphasically on firing 
rate $f$. This is a non-intuitive result, following from the fact that
$P_{ATP}$ is a product of two terms: $I_{p}$ and $(3V_{Na}-2V_{K}-V_{o})$.
The first of them increases monotonically with $f$, whereas the second
decreases with $f$ because $V_{Na}$ and $V_{K}$ depend on frequency
via [Na]$_{av}$. This biphasic dependence of $P_{ATP}$ on frequency has 
interesting implications for thermal properties of brain tissue, which 
are discussed in the subsection (3.7). From the estimated firing rates
for several mammalian species we can also estimate, based on Eq. (19),
their $P_{ATP}$ rates (Table 2).

The sodium pump power $P_{ATP}$ also depends inversely
on the fiber diameter $d$ (Eq. (19) and Fig. 6B), and proportionally
on the gray matter volume $U_{g}$. Thus, too thin fibers are metabolically
expensive.

The power generated by the sodium pumps in the gray matter can be related
directly to the glucose cerebral metabolic rate CMR$_{glu}$ if we
combine Eqs. (16) and (18). The resulting relationship is:

\begin{equation}
P_{ATP}\approx 31 F U_{g}\left(3V_{Na}-2V_{K}-V_{o}\right)\mbox{CMR}_{glu}. 
\end{equation}\\
Thus, the power generated scales linearly with the glucose consumption 
rate. However, it should be kept in mind that the proportionality factor 
is not a constant, but changes with firing rate.

\subsection{Efficiency of the Na$^{+}$/K$^{+}$ pump.}

Let us estimate the efficiency of the sodium pump, i.e., how much
energy does it use for pumping out 3 Na$^{+}$ and pumping in 2 K$^{+}$ 
ions given an available energy from ATP hydrolysis. Energy from hydrolysis 
of 1 ATP molecule goes for performing the useful work of 
$-3e(V-V_{Na}) + 2e(V-V_{K})$, or equivalently 1 mole of ATP performs 
the work of $F(3V_{Na}-2V_{K}-V_{o})$, where $e$ is the electron charge. 
On the other hand, hydrolysis of 1 mole of ATP generates $J_{ATP}$ of 
free energy. Thus, the efficiency of the process is given by 
$F(3V_{Na}-2V_{K}-V_{o})/J_{ATP}$. The value of $J_{ATP}$ depends to
some extent on the internal chemical (ionic) state of the cell, and it
has been reported to be in the range from 48 kJ/mol (Jansen et al, 2003) 
to 62 kJ/mol (Erecinska and Silver, 1989). This leads to the pump 
efficiency of $73-95$ $\%$ (for typical resting values of voltages:
$V_{Na}= 0.068$ V, $V_{K}= -0.100$ V, $V_{o}= -0.067$ V; Hille (2001)).
Because of the high efficiency of the Na$^{+}$/K$^{+}$-ATPase, in what
follows, we make an approximation in which we equate the heat released 
in the gray matter due to hydrolysis of ATP ($G_{ATP}$ in Eq. 9) with the 
electrical power dissipated by the pump ($P_{ATP}$ in Eq. 7).

We can estimate the heat rate for the gray matter of human brain from 
Eq. (20). Taking $U_{g}= 680$ cm$^{3}$ (Stephan et al, 1981), 
$CMR_{glu}= 5.7\cdot 10^{-9}$ mol/(cm$^{3}\cdot$sec) 
(or 0.34 $\mu$mol/(cm$^{3}\cdot$min); e.g. Clarke and Sokoloff, 1994), 
and for the above values of voltages we obtain $P_{ATP}\approx 5.5$ 
Watts. Given a possible increase of this heat value by up to 27$\%$ due to 
pump efficiency, this result does not differ much from other estimates 
of heat in gray matter. For example, Aiello and Wheeler (1995) used mass 
specific heat generation of 11.2 W/kg (based on older experimental data), 
which yields 7.8 Watts for the human gray matter.

\subsection{Scaling of cerebral blood flow with brain size.}

Cerebral blood flow CBF is important in controlling brain temperature
(see the next subsection). The dependence of physiologically averaged
CBF on brain volume can be found from the empirical data available
in the literature. The results in Fig. 7 for 6 mammals spanning 3 orders
of magnitude in brain size show that CBF scales systematically with
brain volume as $\mbox{CBF}= 0.018 U_{br}^{-0.10}$ 1/sec. This implies 
that cerebral blood flow decreases weakly as brains increases in size,
ranging from $\sim 0.020$ sec$^{-1}$ for mouse to
$\sim 0.009$ sec$^{-1}$ for human.

\subsection{Brain temperature vs. frequency and fiber diameter, and
efficiency of brain cooling}

The spatial distribution of brain temperature $T(r)$ is found by solving 
Eq. (9) with $G_{ATP}= P_{ATP}$, and the result is given by (see Appendix C):

\begin{eqnarray}
T(r)= T_{bl} + \frac{P_{ATP}}{\rho_{bl}c_{bl}U_{br}\mbox{CBF}}
- \frac{\left(\sigma_{SB}(T_{sc}^{4}-T_{o}^{4}) + \eta(T_{sc}-T_{o})\right)}
{(\kappa\rho_{bl}c_{bl}\mbox{CBF})^{1/2}} e^{-\xi(R-r)},
\end{eqnarray}\\
where $\xi= (\rho_{bl}c_{bl}\mbox{CBF}/\kappa)^{1/2}$. The parameter $\xi$
characterizes the inverse of the length of ``transition'' region from the 
scalp to the brain's interior where temperature is inhomogeneous (Fig. 8A). 
For human $\xi^{-1}\approx 0.4$ cm, and it is much smaller than human
brain radius ($\approx 8$ cm), and thus heterogeneity is present only
at the brain's edge. On the contrary, for mouse $\xi^{-1}\approx 0.25$
cm, i.e., it is comparable with the mouse brain radius ($\approx 0.4$ cm),
implying that in very small brains cerebral temperature is inhomogeneous
in the whole volume (Fig. 8A). The scalp temperature $T_{sc}$ present
in Eq. (21) is determined self-consistently (see Appendix C) from the
condition $T_{sc}= T(R)$, and it decreases very weakly with brain
volume (Table 2). Its value for human, $T_{sc}= 34.7 ^{o}$C, is similar to 
experimental values ($34-35 ^{o}$C; Hensel et al, (1973)).

Deep brain temperature $T(0)$ is slightly larger than the blood temperature
by $0.1-0.2$ $^{o}$C for all analyzed mammals, except mouse (Table 2).
This value lies in the range of values observed experimentally  
(Hayward and Baker, 1968; Nybo et al, 2002; Kiyatkin, 2007). 
In general, however, $T(0)$ is very weakly species specific. For
large enough brains $T(0)$ is approximately 
$T(0)\approx T_{bl} + P_{ATP}/(\rho_{bl}c_{bl}U_{br}\mbox{CBF})$, 
which shows that
the boundary temperature (and scalp cooling) becomes unimportant in
this limit. From this formula it follows that the deep brain temperature
is greater than the blood (or core body) temperature by the quantity
proportional to the ratio of CMR$_{glu}$ to CBF, in agreement with
Yablonskiy et al (2000). This result suggests for example that, for 
a given animal, any local increase in the cerebral blood flow that 
exceeds an increase in glucose utilization rate leads to a lowering 
of brain temperature, and vice versa.

Deep brain temperature $T(0)$ depends significantly on the effective
fiber diameter $d$ (Fig. 8B,C). For very small $d$ (corresponding to very
small $d_{a}$), the temperature $T(0)$ tends to diverge,
which is a direct consequence of the fact that $P_{ATP}$ also diverges 
for $d\rightarrow 0$ (or equivalently $d_{a}\rightarrow 0$).
In this ``thin fiber'' limit we have a simple inverse proportionality 
(see Eqs. (19) and (21)) between brain-body temperature difference and 
fiber thickness, i.e., $T(0)-T_{bl} \sim 1/d \sim 1/(2d_{a})$.
This relationship implies that decreasing fiber diameter twofold
increases $T(0)-T_{bl}$ by the same amount. This suggests again that 
too thin fibers are not beneficial for brain thermal equilibrium and 
hence its functioning.

In contrast to the monotonic dependency of $T(0)$ on $d$, its dependence 
on firing rate is biphasic (Fig. 8D). 
The origin of this non-monotonic behavior is the dependence
of $P_{ATP}$ on $f$, which is also biphasic (see Fig. 6A). This result
has surprising thermodynamic consequences, namely too high levels of
neural activity (firing rate) can lead to a decrease in the cerebral
tissue temperature. For human brain, corresponding changes in temperature 
are rather small, at the peak about 0.5-0.6 K (Fig. 8D), which falls
into the range of values reported experimentally (Yablonskiy et al, 2000;
Kiyatkin, 2007).

In the superficial regions brain temperature is always smaller than the
blood temperature (Fig. 8A; Table 2). This fact has important consequences
for the heat transfer in the mammalian brains. In deep brain regions
cerebral blood flow plays the role of a coolant, whereas in the superficial
regions it serves as a strong brain heater. Consequently, the net effect
of blood flow is to warm up the brain tissue, i.e., $\dot{Q}_{bl} < 0$
(Table 2).

In general, the cooling mechanisms discussed in subsection (2.5) depend
on brain temperature distribution, on scalp temperature $T_{sc}$, and
on brain volume $U_{br}$. Therefore, the efficiency of cerebral cooling
is species specific. In particular, the relative importance of the two
major mechanisms of heat transfer inside the brain is dependent on brain 
size. For very small brains, heat cooling rate via conduction $\dot{Q}_{c}$
is comparable to the heat warming rate via blood flow $\dot{Q}_{bl}$
(Table 2). The reason for such a strong warming through the cerebral 
blood is that it must compensate heat loss due to scalp convection/conduction
$\dot{Q}_{cv}$ and radiation $\dot{Q}_{r}$, which are much larger than the
metabolic rate $P_{ATP}$. On the other hand, for large brains, the cooling
rate $\dot{Q}_{c}$ is almost twice as large as the warming rate 
$\dot{Q}_{bl}$, because heat production due to metabolic activity $P_{ATP}$
is more significant than for small brains (Table 2). On the scalp, 
the heat transfer rate is dominated by convection/conduction $\dot{Q}_{cv}$,
as it is twice the radiation rate $\dot{Q}_{r}$.

\subsection{Thermal bounds on fiber diameter and length.}

Mammals are able to sustain brain temperatures up to about 42 $^{o}$C 
without causing brain damage (Gordon, 1993; Kiyatkin, 2007). 
Above these temperatures molecular changes in neurons and synapses 
become critical and irreversible. In what follows, we estimate the 
thermal bounds on fiber diameter that allow to maintain the safe 
temperature regime. For this purpose we use Eq. (21) for $r\approx 0$, 
in which we put $\Delta T_{max}\equiv T(0) - T_{bl}= 5$ K, 
as the largest possible difference between brain and blood
temperatures. Solving Eq. (21) for the effective fiber diameter $d$, 
we obtain

\begin{eqnarray}
d \ge d_{min}= 
\frac{4(1-\phi)U_{g}\left(g_{Na,o} + f(C+\delta C)\right)}
{\frac{3\rho_{bl}c_{bl}U_{br}\mbox{CBF} \Delta T_{max}}
{(V_{Na}-V_{o})(3V_{Na}-2V_{K}-V_{o})} 
- \frac{fq\rho_{s}g_{s}\tau_{s}V_{K}U_{g}}{(V_{K}-V_{Na})} },
\end{eqnarray}\\
where we used Eq. (19) for $P_{ATP}$. The minimal effective fiber 
diameter $d_{min}$ depends very weakly on gray matter
volume $U_{g}$ (Fig. 9A). This suggests that brain size is not 
a critical factor determining thermodynamic safety of the cerebral 
tissue. In other words, thermal properties of mouse and elephant brains,
differing by 4 orders of magnitude in size, are rather similar.

The minimal effective fiber diameter $d_{min}$ depends stronger and 
biphasically on firing rate $f$ (via $V_{Na}$, $V_{K}$, and [Na]$_{av}$) 
(Fig. 9B). For very low firing rates ($f\rightarrow 0$), the expression 
(22) simplifies, and we obtain in this limit:

\begin{eqnarray}
d_{min}\approx \frac{4(1-\phi)g_{Na,o}(V_{Na}-V_{o})
(3V_{Na}-2V_{K}-V_{o})U_{g}}
{3\rho_{bl}c_{bl}U_{br}\mbox{CBF} \Delta T_{max}},
\end{eqnarray}\\
which yields $\approx 0.3$ nm (values of $U_{g}$ and $U_{br}$ are for 
human brain). For intermediate values of $f$, for
which $d_{min}$ has a maximum, the value of $d_{min}$ can be in the range
$0.04-0.05$ $\mu$m (Fig. 9B), which corresponds to a bound on axon
diameter $d_{a}= 0.02-0.03$ $\mu$m. The latter value is only about 5 times 
larger than the membrane thickness (Koch, 1998). The term proportional to 
$fq$ in the denominator of Eq. (22) is the background synaptic contribution 
with frequency dependent depression inside $q$. Due to this depression the 
synaptic contribution is bounded from above and consequently does not 
diverge as a function of $f$. Thus, synaptic depression not only reduces 
the power $P_{ATP}$ (see. Eq. 19), but it also makes $d_{min}$ finite.
Without depression, the bound $d_{min}$ would diverge for $f\sim 450$ Hz.

Average value of the empirical fiber diameter $d$ in the gray matter
is brain size independent and is about $0.45$ $\mu$m (harmonic mean of 
average axon and dendrite diameters in mouse, as defined in Sec. 2.2, 
Eq. (3)). This value is about 10 times larger than the largest value of 
$d_{min}$. Because of this, mammalian brains operating under normal 
physiological conditions are rather safe from excessive overheating 
that would cause brain damage, for all ranges of frequency. 
The situation could be more tricky in hot environments in which 
$\Delta T_{max}$ were severely reduced, due to increase in body 
temperature and hence cerebral blood temperature.
The margins of thermal safety, could be also compromised in pathological 
conditions. For example, abnormalities associated with strongly reduced 
cerebral blood flow CBF or compromised synaptic depression, could 
significantly increase the bound $d_{min}$ up to the lower neuroanatomical 
values of $d$.

The thermal lower bound on fiber diameter also determines the upper
bound on fiber length per neuron $l$, as is evident from Eq. (4). Taking
the neuron's surface area $S=\pi ld$, we get 
$l \le 4(1-\phi)U_{g}/(\pi Nd_{min}^{2})$, where $d_{min}= 0.01-0.05$
$\mu$m. For human brain with $U_{g}= 680$ cm$^{3}$ (Stephan et al, 1981)
and $N= 2\cdot 10^{10}$ (Haug, 1987; Braendgaard et al, 1990) this
yields $l \le 12-300$ meters. For mouse brain with $U_{g}=0.11$ cm$^{3}$
and $N= 1.6\cdot 10^{7}$ (Braitenberg and Schuz, 1998) we get 
$l \le 2.4-60$ meters. Neuroanatomical data for mouse gray matter 
indicate that the average fiber length per neuron is $1.5-4.5$ cm
(Braitenberg and Schuz, 1998), which is 2-3 orders of magnitude below
its upper thermal bound. A similar conclusion holds for the human brain.

Taken together all these results suggest that thermodynamics does not
restrict neuroanatomical parameters in any dramatic way, because they 
are far away from their thermal bounds.

\newpage

\section{Discussion}

This article investigates thermodynamic properties of brain tissue
and corresponding physical limits on neural anatomy caused by heat 
balance in the brain. It is found that, in general, the lower and upper 
limits on fiber dimensions are unattainable for normal values of 
physiological parameters such as cerebral blood flow and maximal 
Na$^{+}$/K$^{+}$-ATPase. This suggests that real mammalian brains either 
keep these physiological parameters in the ``proper'' range or scale 
appropriately average fiber diameter and length to maintain wide margins 
of thermodynamic safety. Such wide margins presumably enable the brain 
to avoid overheating during enhanced cerebral activities, abnormal states 
(e.g. epileptic seizures), and in hot environments.

The conclusions of this paper are based on calculating neural
metabolic power and spatial distribution of heat dissipated. Since
majority of metabolic energy in neurons goes to pumping out sodium ions 
(Ames, 2000; Astrup et al, 1981; Erecinska and Silver, 1989; 
Rolfe and Brown, 1997), the first step in determining neural power
was to find Na$^{+}$ influx during an action potential. An explicit
analytical formula for this Na$^{+}$ influx was derived (Eq. 12)
and was validated numerically. This analytical formula yields similar
results for an isolated action potential as the one of phenomenological
character by Attwell and Laughlin (2001) ($20-28 \%$ discrepancy
between the two can be attributed to the different assumptions regarding
the magnitude of depolarizations in axons and dendrites). However,
for repetitive firing the derived here Eq. (12) extends the Attwell
and Laughlin (2001) formula, because it accounts for intracellular
sodium accumulation and corresponding changes in membrane electrical
properties expressed by $\delta C$ (correction to Na$^{+}$ influx
due to prolonged Na$^{+}$ channels activation). Moreover, the sodium
build-up is also important for the accurate determination of ATP/glucose
utilization rates for high frequencies. In this respect, Eqs. (16) and 
(17) extend the Attwell and Laughlin (2001), and Lennie (2003) linear
calculations to the non-linear regime. An additional novelty in the
present approach is inclusion of synaptic depression in the transmission
probability.

The issue of brain cooling is not a classic problem of volume to
surface ratio, as it is the case with natural, non-designed, physical
objects. Instead, the brain cooling could be compared to the cooling 
of combustion heat engine, which receives a liquid coolant. In the brain the 
role of the coolant is played by the cerebral blood, but only in the deep
region because there blood has a slightly lower temperature than the 
brain tissue. In the superficial regions brain tissue has a smaller 
temperature than the cerebral blood, and there blood warms up the brain. 
The fact that the deep brain temperature depends weakly on brain volume 
(Table 2), implies that brain size is not a major determinant of 
thermal responses. This in turn implies
that the thermodynamics of heat balance does not restrict the brain
size in any significant way, suggesting that, in principle, brains
could be heavier than 5 kg (the largest known brain).

The interesting result is that the power generated by the sodium pump 
depends biphasically on neural frequency of firing, and inversely on the 
effective fiber diameter (Fig. 6). As a consequence of this, brain temperature
$T(0)$ can depend biphasically on frequency as well, if cerebral blood 
flow does not change with frequency (Fig. 8D). Thus $T(0)$ increases
with frequency but only up to a certain point above which it slightly
decreases with further increase in frequency. The increase of brain 
temperature in response to activation (higher firing rate) is an expected 
behavior (Kiyatkin, 2007). However, the decrease of $T(0)$ for very
large stimulation (very high firing rates) is an unexpected effect that 
can explain some experimental results in which stimulation of certain 
brain regions led to lowering of local temperature (McElligott and Melzack, 
1967; Yablonskiy et al, 2000). The standard explanation for the decrease 
of $T(0)$ upon stimulation requires that cerebral blood flow must increase
more than $P_{ATP}$ increases (Yablonskiy et al, 2000). The biphasic
relationship between temperature and frequency, obtained in this paper 
(Fig. 8D), offers an alternative explanation, namely, that brain temperature 
can decrease upon vigorous (or prolonged) stimulation even when cerebral 
blood flow does not change.

The negative correlation between the power generated in the brain and the
effective wire diameter (Fig. 6B) suggests that thin fibers are energy 
expensive. The effective fiber diameter $d$ is defined as a harmonic mean
of axon $d_{a}$ and dendrite $d_{d}$ diameters (Sec. 2.2), with the
assumption that volumes of axons and dendrites are equal (Braitenberg and
Sc{\"u}z, 1998; Chklovskii et al, 2002). In general, $d$ and $d_{a}$ are
positively correlated and $d\rightarrow 0$ if $d_{a}\rightarrow 0$.
Thin fibers also promote large fluctuations of intracellular sodium 
concentration in response to stochastic action potentials (Fig. 3B), 
which can be disadvantageous for electrical properties of neurons.

Another negative consequence of having too thin fibers is that cerebral 
tissue temperature increases inversely with fiber diameter (Fig. 8C),
i.e., $T(0)-T_{bl} \sim 1/d$. Thus, brain regions reach in very thin
wire can heat up excessively. As an example, the temperature of the
thinnest known axons with diameter 0.1 $\mu$m (Faisal et al, 2005, i.e. 
$d= 0.18$ $\mu$m), relative to blood temperature, should be about 2.5 times 
larger than the corresponding relative temperature of 0.3 $\mu$m axons 
(mouse cerebral cortex; Braitenberg and Schuz, 1998; corresponding to 
$d=0.45$ $\mu$m), assuming both axons have the same firing rate. Thus,
if typical relative temperatures above blood temperature are $0.2-0.3$ $^{o}$C
for 0.3 $mu$m axons (Hayward and Baker, 1968), then for 0.1 $\mu$m axons
the corresponding relative temperatures would be $0.5-0.8$ $^{o}$C.
This result suggests an explanation why in the peripheral nervous system 
sensory fibers responsible for high-threshold heat sensation 
(so-called C-fibers) are much thinner ($0.1-1.5$ $\mu$m;
Kandel et al, 1991) than other sensory fibers (A-type or B-type with
diameters $1-5$ $\mu$m). It might be that they warm up easier, although
these fibers also serve other functions (Craig, 2003).

This study finds that the lower thermal bound on the effective fiber 
diameter $d_{min}$ is strongly frequency dependent, but it is finite due 
to synaptic depression. Values of $d_{min}$ are in the range $0.3-50$ nm 
(or $0.0003-0.05$ $\mu$m; see Fig. 9B). These values can be translated
to the corresponding thermal bounds on axon diameter $d_{a}$, which are
in the range $0.0002-0.026$ $\mu$m (assuming that the dendrite diameter
$d_{d}= 0.9$ $\mu$m and it does not change). The average value of cortical 
axon diameter is $d_{a}=0.3$ $\mu$m, and thus it is 12-1500 times larger 
than these limits, therefore, on average, mammalian brains operate in the 
safe thermal zone. However, it should be also kept in mind that axon diameter 
displays some variability, and the thinnest axons can reach 0.1 $\mu$m 
(Faisal et al, 2005). This value is still about 4 times larger than the 
maximal value of the lower thermal limit on $d_{a}$ (which is 0.026 $\mu$m), 
suggesting that thermal limits are not an immediate constraint on fiber size.
It is also interesting to note that the estimated maximal lower thermal 
bound on the diameter of brain axonal wiring ($0.026$ $\mu$m) is smaller than
corresponding bounds imposed by structural constraints and noise, which 
are respectively $0.06$ $\mu$m and $0.1$ $\mu$m (Faisal et al, 2005).

The theoretical bound on axon diameter implies a corresponding lower 
bound on the speed of signal propagation in the gray matter. Let us 
estimate the upper limit on temporal delays. For unmyelinated fibers 
(prevalent in gray matter) velocity of signal propagation is proportional 
to the square root of the fiber diameter (Hodgkin, 1954). Experimentally, 
for axons with $d= 1$ $\mu$m the propagation velocity is 2.3 mm/msec 
(Koch, 1998), which implies that for axons with the boundary thickness
of $0.026$ $\mu$m, we have velocities 0.37 mm/msec.
This gives for the maximal known extent of axons in the gray matter of 
9-10 mm (for macaque monkey visual cortex; Amir et al, (1993))
the upper limit on delays in the range 24-27 msec. Thus, apparently, 
thermodynamics of heat balance in the gray matter does not tolerate 
temporal axonal delays longer than $\sim$ 0.03 sec, which is a stringent
constraint.

The problem of finding bounds on fiber dimensions is similar in spirit
to the approaches of ``wire minimization'' in the brain (Cherniak 1995; 
Murre and Sturdy 1995; Karbowski 2001, 2003; Chklovskii et al, 2002).
It is hypothesized that this principle governs the organization of the 
mammalian nervous system at different scales 
(Murre and Sturdy 1995; Prothero, 1997; Kaas 2000; Karbowski 2001, 2003),
because it offers energy savings associated with ionic membrane transport,
as well as reduction in temporal delays in neural communication. The heat 
balance limits on fiber diameter, found in this paper, suggest corresponding
upper thermal bounds on the density of wire packing in the brain in
the range $20-100$ fibers per $\mu$m.
These thermal bounds on wire density are a factor of 2-4 higher
than the corresponding upper bounds coming from structural and noise 
considerations (Faisal, 2005).

The estimated firing rates in mammals are in the range from 1.7 Hz for
human to 6.2 Hz for mouse, and they scale systematically with brain size, 
with the exponent $-0.15$ (Fig. 5C). The estimate for rat (5 Hz) is very 
close to that assumed by Attwell and Laughlin (2001), i.e. 4 Hz, which was 
based on weighted average of values observed experimentally. Also the 
estimate for cat (4.5 Hz) is reasonably close to that reported 
experimentally for visual cortex (2.5-4.0 Hz; see Baddeley et al, 1997). 
However, {\it in vivo} values of firing rates are stimulus dependent and 
in many species are largely unknown. The scaling result of firing rates 
is qualitatively consistent with the experimental data on avian brains, 
which show an allometric decay of firing rates in the peripheral nervous
system with brain/body mass (Hempleman et al, 2005). It is also consistent 
with a prediction coming from a recent analysis of empirical data on 
brain metabolic scaling in mammals (Karbowski, 2007), where it was 
suggested that an average firing rate should decrease for bigger brains 
to account for a negative allometric exponent of specific metabolic rate,
which was also $-0.15$. 
This conclusion is analogous to the general trend in mammals, in which 
physiological processes tend to slow down with an increase in body size 
(Schmidt-Nielsen 1984).

The estimate of the heat generated in the gray matter may have some 
margins of error. First, the assumption of the equipotential neuron
is only an approximation, because it does not include explicitly the
spatio-temporal effects associated with action potential propagation
and back-propagation. However, it is estimated in Appendix A that
the correction from this effect is rather small of the order of 16$\%$,
i.e., the actual Na$^{+}$ influx can be larger by this amount. Second,
because the efficiency of the pump is a little less than 100$\%$
($73-95$ $\%$; Sec. 3.5), the actual heat dissipated in the gray matter 
may be larger than $P_{ATP}$ by $5-27$ $\%$. Therefore the total error 
from these two effects on the cerebral heat can theoretically reach 43$\%$.
We should also remember that some heat coming from the glucose to ATP
conversion was neglected (i.e., it is assumed that glycolysis is 100$\%$
efficient). However, the comparison of the heat estimates for the human 
gray matter indicates that in fact the total heat error cannot be larger 
that 30$\%$ (Sec. 3.5). There are also other sources of error that affect 
the temperature distribution in the brain, such as the neglect of cooling 
by scalp perspiration, and variability of environmental $T_{o}$ temperature, 
which can affect convection/conduction and radiation cooling rates. 
However, these contributions should not have any dramatic influence
on the thermal properties of brain tissue. Moreover, neither of these 
sources of error or their combination seem to affect the main conclusion 
of this paper, namely, that neural anatomy (fiber diameter and its length) 
does not approach its thermodynamic limits. To reach these limits the ratio 
of the maximal sodium pump current to the cerebral blood flow would have 
to be at least 10-100 times larger.

The formulas (Eqs. 17-21) in this paper for the CMR$_{glu}$, power 
dissipated in the gray matter, and brain temperature, may have practical 
use. These formulas as well as their possible future extensions can be 
used for assessing neural activity (firing rates) based on changes in 
temperature, CMR$_{glu}$, and CBF. Moreover, because this study offers 
a direct relationship between neural activity and neuroanatomical parameters, 
it may be useful in quantitative studies of the interplay between brain 
development, evolution, and metabolism (Purves, 1988; Striedter, 2005).

\vspace{0.5cm}

\noindent{\bf Acknowledgments}

The work was partly supported by the Caltech Center for Biological 
Circuit Design. I acknowledge useful suggestions of the two anonymous 
reviewers.

\vspace{1.5cm}

\noindent{\bf Appendix A: Sodium influx during an action potential.}

The duration of a typical action potential can be divided into two
phases (Fig. 2A). During the first phase Na$^{+}$ conductance $g_{Na}$ 
rises almost instantenously to its maximal value $\overline{g}_{Na}$ and
voltage $V$ increases to its peak value $V_{Na}$. The second phase is
characterized by decline in values of $g_{Na}$ (to zero) and $V$ 
(to values $< 0$). The total Na$^{+}$ influx during an action potential
is $\Delta[\mbox{Na}]_{o}= \Delta[\mbox{Na}]_{o}^{(1)} 
+ \Delta[\mbox{Na}]_{o}^{(2)}$,
where superscripts (1) and (2) refer to the first and second phase,
respectively.

During the first phase, the intracellular potassium concentration 
practically does not change, because K$^{+}$ channels are activated
with a delay. Thus, by straightforward integration of Eq. (2), we
obtain the Na$^{+}$ influx during this phase as:

\begin{eqnarray}
\Delta[\mbox{Na}]_{o}^{(1)}\approx  \frac{4C}{Fd}(V_{Na}-V_{o}),
\end{eqnarray}\\
where the contribution proportional to $g_{L}$ was neglected, since
it is much smaller.

Sodium influx during the second phase can be computed from the sodium
dynamics of Eq. (6). Most of the time during this phase, the term
proportional to $g_{Na}$ is much larger than the remaining two terms
in Eq. (6). Thus, we can write

\begin{eqnarray}
\Delta[\mbox{Na}]_{o}^{(2)}\approx - \frac{4}{Fd} \int_{0}^{\tau_{o}} dt \;
g_{Na}(t)(V(t)-V_{Na}),
\end{eqnarray}\\
where $\tau_{o}$ is the duration of the second phase and 
$\tau_{o}\approx 0.4$ msec (from simulations). To simplify calculations,
it is assumed that Na$^{+}$ conductance and voltage depend on time 
in the following way: $g_{Na}(t)\approx \overline{g}_{Na}(1-t/\tau_{o})$
and $V(t)\approx V_{Na} +  (0.6V_{K}-V_{Na})(t/\tau_{o})^{z}$.
These forms assure that for $t= 0$, we have 
$g_{Na}(0)\approx \overline{g}_{Na}$ and $V(0)\approx V_{Na}$, and
for $t= \tau_{o}$ we have 
$g_{Na}(\tau_{o})\approx 0$ and $V(\tau_{o})\approx 0.6V_{K}$.
The latter value comes from simulations. After performing integral
in Eq. (25) we obtain:

\begin{eqnarray}
\Delta[\mbox{Na}]_{o}^{(2)}\approx  \frac{4\overline{g}_{Na}\tau_{o}
(V_{Na}-0.6V_{K})}{Fd(z+1)(z+2)}
\end{eqnarray}\\
The fits in Fig. 4 are made for $z=2.5$. The total Na$^{+}$ influx
during an action potential is $\Delta[\mbox{Na}]_{o}^{(1)} 
+ \Delta[\mbox{Na}]_{o}^{(2)}$, 
and is given by Eq. (12) in the main text.

It is also interesting to check the magnitude of correction coming from
the fact that real neurons are not equipotential and action potentials
propagate and back-propagate with a finite velocity. Let $\zeta$ be 
the spatial constant characterizing membrane potential homogeneity, and
$c$ the velocity of action potential propagation (along axon) and 
back-propagation (along dendrites). A simple way to account for the
spatio-temporal dependence of Na$^{+}$ conductance and voltage in Eq. (25)
is to make the following rescalings: 

$g_{Na}(x,t)\approx \overline{g}_{Na}
(1 - (t-x/c)/\tau_{o}) \exp(-(ct-x)/\zeta) H(ct-x)H(x+c\tau_{o}-ct) $

and

$V(x,t)\approx \{ V_{Na}(1 - (t-x/c)^{z}/\tau_{o}^{z}) \exp(-(ct-x)/\zeta) 
 + 0.6V_{K}(t-x/c)^{z}/\tau_{o}^{z} \}H(ct-x)H(x+c\tau_{o}-ct) $,

where the factor $\exp(-(ct-x)/\zeta)$ denotes a traveling wave of excitation 
along axon or dendrite with the spatial spread $\zeta$, and the function 
$H(y)$ is the standard Heaviside function equal to 1 for $y \ge 0$ and equal 
to 0 for $y < 0$. These rescalings assure that points along axon and dendrite 
separated by $x$ from the soma (action potential initiation zone) receive 
an excitation after the delay time $x/c$. With these modifications, after 
some algebra, the corrected result for the $\Delta[\mbox{Na}]_{o}^{(2)}$ is

\begin{eqnarray}
\Delta[\mbox{Na}]_{o}^{(2)}\approx  \frac{4\overline{g}_{Na}\tau_{o}
(V_{Na}-0.6V_{K})}{Fd(z+1)(z+2)} \left( 1 + \epsilon \frac{c\tau_{o}}{\zeta}
\right),
\end{eqnarray}\\
where $\epsilon$ is given by $\epsilon= \frac{(z+1)}{(V_{Na}-0.6V_{K})}  
\{ (z+2)V_{Na}/6 - (2V_{Na}-0.6V_{K})/(z+3) \}$. 
The term proportional to $c\tau_{o}/\zeta$ represents the first order 
correction and it is clear that the equipotential approximation
corresponds to the case when $\zeta= \infty$ or $c\tau_{o}/\zeta \ll 1$. 
For physiological values of parameters: 
$V_{Na}= 60$ mV, $V_{K}= -100$ mV, $c= 2.3$ mm/msec, $\zeta= 1$
mm (Koch, 1998), with $\tau_{o}= 0.4$ msec and $z=2.5$, we obtain
$\epsilon c\tau_{o}/\zeta \approx 0.22$. This implies that the 
equipotential assumption slightly underestimates the sodium influx during 
the second phase of sodium activation. Overall, this correction to the 
total Na$^{+}$ influx $\Delta[\mbox{Na}]_{o}$ is small, $\sim 16\%$, 
for low firing rates for which $\delta C\approx 2.4C$, and it gets even 
smaller for high firing rates. A detailed numerical treatment of the influence 
of action potential velocity on metabolic rate of the squid giant axon 
is presented in (Crotty et al, 2006).

\vspace{0.7cm}

\noindent{\bf Appendix B: Relaxation time constant for Na$^{+}$ dynamics.}

In this Appendix the relaxation process of voltage and sodium to their
equilibrium values is analyzed following a single action potential
for a neuron that prior to that has been at rest for a long time.
Computations below are performed in a very late phase of an action
potential when the conductances $g_{Na}$, $g_{K}$, and the voltage $V$ 
returned essentially to their resting values 
(or are close to them, i.e. $dV/dt\approx 0$).
During this phase we can use a linear approximation on $V$ and [Na]
in Eq. (6). That is, we can expand $V$ and $[Na]$ around their resting values 
$V_{o}$ and $[\mbox{Na}]_{o}$ as: $V\approx V_{o} + \Delta V$, 
$[\mbox{Na}]\approx [\mbox{Na}]_{o} + \Delta[\mbox{Na}]$, 
where  $\Delta[\mbox{Na}]/[\mbox{Na}]_{o} \ll 1$,
and neglect higher order terms.
In this approximation the pump current $I_{p}$ can be written
as $I_{p}\approx I_{p,o} + \lambda\Delta[\mbox{Na}]$, where
$I_{p,o}= AS [\mbox{Na}]^{k}_{o}/([\mbox{Na}]^{k}_{o} + \theta^{k})$, and
$\lambda= kA\theta^{k}[\mbox{Na}]^{k-1}_{o}/(\theta^{k}+[\mbox{Na}]^{k}_{o})^{2}$.
The corresponding changes in $V_{Na}$ and $V_{K}$ are
$\Delta V_{Na}\approx -(RT/F)\Delta[\mbox{Na}]/[\mbox{Na}]_{o}$ and
$\Delta V_{K}\approx -(RT/F)\Delta[\mbox{K}]/[\mbox{K}]_{o}= 
(RT/F)\Delta[\mbox{Na}]/[\mbox{K}]_{o}$.
Next, using the facts that $g_{Na}/g_{L} \ll 1$ and $g_{K}/g_{L} \ll 1$, 
we can solve Eq. (6) analytically. The equations governing $\Delta[\mbox{Na}]$
and $\Delta V$ relaxations are given by:

\begin{eqnarray}
\frac{dF}{4}\frac{d}{dt}\Delta[\mbox{Na}]= 
- \left\{3\lambda + \frac{RT}{F[\mbox{Na}]_{o}}
\left(g_{Na} - \frac{qf\rho_{s}dg_{s}\tau_{s}V_{K}(V_{o}-V_{K})}
  {4(1-\phi)(V_{K}-V_{Na})^{2}}\right) \right\}\Delta[\mbox{Na}]
\nonumber \\
\Delta V(t)\approx - \frac{\left\{\lambda + (RT/F)(g_{Na}/[\mbox{Na}]-
g_{K}/[\mbox{K}])
\right\}
\Delta[\mbox{Na}]}{\left\{g_{L} 
+ qf\rho_{s}g_{s}\tau_{s}d/(4(1-\phi))\right\}}.
\end{eqnarray}\\
The term on the right hand side of $\Delta[\mbox{Na}]$ dynamics that is 
proportional to $RT/(F[\mbox{Na}]_{o})$ is much smaller than $3\lambda$,
and thus it can be neglected. This implies that changes in $V_{Na}$ and
$V_{K}$ due to sodium influx for an isolated action potential are very
small. This leads to a simple exponential decay of $\Delta[\mbox{Na}]$ as 
$\Delta[\mbox{Na}]\approx \Delta[\mbox{Na}]_{o} e^{-t/\tau}$ with the time 
constant $\tau\approx Fd/(12\lambda)$, where $\Delta[\mbox{Na}]_{o}$ is 
Na$^{+}$ influx
during the action potential given by Eq. (12). Thus, relaxation time
constant is short for thin fibers, and it increases proportionally with
fiber diameter. As an example, for $d= 0.45$ $\mu$m we obtain 
$\tau\approx 5$ sec (for $k=3$ and $[\mbox{Na}]_{o}=12.0$), which is of the
right order of magnitude (Abercrombie and Weer, 1978; Nakao and Gadsby,
1989). For $d=0.18$ $\mu$m the time constant $\tau$ is about 2 sec.

\vspace{0.7cm}

\noindent{\bf Appendix C: Solution of the thermal balance equation.}

The steady-state limit of Eq. (9) can be rewritten in the form:

\begin{eqnarray}
\frac{\partial^{2} T}{\partial r^{2}}= \frac{\rho_{bl}c_{bl}\mbox{CBF}}
{\kappa}(T-T_{o}), 
\end{eqnarray}\\
where $T_{o}= T_{bl} + G_{ATP}/(\rho_{bl}c_{bl}\mbox{CBF} U_{br})$.
In computations, we assume $G_{ATP}\approx P_{ATP}$ because of the high
efficiency of Na$^{+}$/K$^{+}$-ATPase (Sec. 3.5).
We look for the solution of Eq. (29) in the form: 
$T(r)= T_{o} + a\exp(\xi r)$. 
Substituting this form in Eq. (29), we obtain
 $\xi= \pm (\rho_{bl}c_{bl}\mbox{CBF}/\kappa)^{1/2}$, 
and we take only the positive solution as the one corresponding to 
the physical situation. From the boundary condition in Eq. (10) 
we obtain the coefficient 
$a= - \left\{\sigma_{SB}(T_{sc}^{4}-T_{o}^{4}) + \eta(T_{sc}-T_{o})\right\}
e^{-\xi R}/(\kappa\xi)$. Combining these results we find Eq. (21)
in the main text. The scalp temperature $T_{sc}$ is determined numerically
from the condition $T_{sc}= T(R)$, for all considered species.

\newpage

\vspace{1.5cm}

\noindent{\large\bf  References} \\
Abercrombie RF, De Weer P (1978) Electric current generated by squid giant
axon sodium pump: external K and internal ADP effects. {\it Am. J. Physiol.}
{\bf 235}: C63-C68. \\
Aiello LC, Wheeler P (1995) The expensive-tissue hypothesis: the
brain and the digestive-system in human and primate evolution.
{\it Curr. Anthrop.} {\bf 36}: 199-221. \\
Ames III A (2000) CNS energy metabolism as related to function.
{\it Brain Research Reviews} {\bf 34}: 42-68. \\
Amir Y, Harel M, Malach R (1993) Cortical hierarchy reflected in the
organization of intrinsic connections in macaque monkey visual cortex.
{\it J. Comp. Neurol.} {\bf 334}: 19-46. \\
Astrup J, Sorensen PM, Sorensen HR (1981)
Oxygen and glucose consumption related to Na$^{+}$-K$^{+}$ transport
in canine brain. {\it Stroke} {\bf 12}: 726-730. \\
Attwell D, Laughlin SB (2001) An energy budget for signaling in the 
gray matter of the brain. {\it J. Cereb. Blood Flow Metabol.} 
{\bf 21}: 1133-1145. \\
Baddeley R, et al (1997) Responses of neurons in primary and inferior
temporal visual cortices to natural scenes. {\it Proc. R. Soc.
Lond.}  {\bf B 264}: 1775-1783. \\
Baker MA (1982) Brain cooling in endotherms in heat and exercise.
{\it Annu. Rev. Physiol.} {\bf 44}: 85-96. \\
Bennett CH (1982) The thermodynamics of computation - a review. 
{\it Int. J. Theor. Physics} {\bf 21}: 905-940. \\
Braendgaard H, et al (1990) The total number of neurons in the human
neocortex unbiasedly estimated using optical disectors. {\it J. Microsc.}
{\bf 157}: 285-304. \\
Braitenberg V, Sch{\"u}z A (1998) {\it Cortex: Statistics
and Geometry of Neuronal Connectivity.} Berlin: Springer. \\
Busija DW (1984) Sympathetic nerves reduce cerebral blood flow during 
hypoxia in awake rabbits. {\it Am. J. Physiol.} {\bf 247}: H446-H451. \\
Cherniak C (1995) Neural component placement. {\it Trends Neurosci.}
{\bf 18}: 522-527. \\
Chklovskii DB, Schikorski T, Stevens CF (2002) Wiring optimization
in cortical circuits. {\it Neuron} {\bf 43}: 341-347. \\
Clarke DD, Sokoloff L, In "Basic Neurochemistry", ed: Siegel GJ et al
(New York, Raven Press, 1994), pp. 645-680. \\
Cragg BG (1967) The density of synapses and neurones in the
motor and visual areas of the cerebral cortex. {\it J. Anatomy\/}
{\bf 101}: 639-654. \\
Craig AD (2003) Interoception: the sense of the physiological condition
of the body. {\it Curr. Opin. Neurobiol.} {\bf 13}: 500-505. \\
Crotty P, Sangrey T, Levy WB (2006) Metabolic energy cost of action
potential velocity. {\it J. Neurophysiol. } {\bf 96}: 1237-1246. \\
Dayan P, Abbott LF (2001). {\it Theoretical Neuroscience}. 
Cambridge, MA: MIT Press. \\
DeFelipe, J., Alonso-Nanclares, L., and Avellano, J (2002)
Microstructure of the neocortex: Comparative aspects. {\it J.
Neurocytology} {\bf 31}: 299-316. \\
Erecinska M, Silver IA (1989) ATP and
brain function. {\it J. Cereb. Blood Flow Metab.} {\bf 9}: 2-19. \\
Faisal AA, White JA, Laughlin SB (2005) Ion-channel noise places
limits on the miniaturization of the brain's wiring. {\it Curr. Biol.}
{\bf 15}: 1143-1149. \\ 
Falk D (1990) Brain evolution in Homo: the ``radiator'' theory.
{\it Behav. Brain. Sci.} {\bf 13}: 333-381. \\
Frietsch T et al (2007) Reduced cerebral blood flow but elevated 
cerebral glucose metabolic rate in erythropoietin overexpressing 
transgenic mice with excessive erythrocytosis. {\it J. Cereb. Blood 
Flow Metabol.} {\bf 27}: 469-476. \\
Gordon CJ (1993)  {\it Temperature regulation in laboratory rodents}.
Cambridge, UK: Cambridge Univ. Press. \\
Hayward JN, Baker MA (1968) Role of cerebral arterial blood temperature
in the regulation of brain temperature in the monkey. {\it Am. J. Physiol.}
{\bf 215}: 389-403. \\
Haug H (1987) Brain sizes, surfaces, and neuronal sizes
of the cortex cerebri: A stereological investigation of Man and his 
variability and a comparison with some mammals (primates, whales,
marsupials, insectivores, and one elephant). {\it Am. J. Anatomy}
{\bf 180}: 126-142. \\
Hempleman SC, et al (2005) Spike firing allometry in
avian intrapulmonary chemoreceptors: matching neural code to body size. 
{\it J. Exp. Biol.} {\bf 208}: 3065-3073. \\
Hensel H, Bruck K, Raths P (1973). {\it Homeothermic organisms.}
In: {\it Temperature and Life.} Edited by: Precht H et al. New York:
Springer-Verlag, p. 509-564. \\
Hille B (2001) {\it Ionic channels of excitable membranes}. Sunderland, 
MA: Sinauer Assoc., 3rd edition.  \\
Hodgkin AL (1954) A note on conduction velocity. {\it J. Physiol.}
{\bf 125}: 221-224. \\ 
Jansen MA et al (2003) Energy requirements for the Na$^{+}$ gradient
in the oxygenated isolated heart: effect of changing the free energy
of ATP hydrolysis. {\it Am. J. Physiol. Heart Circ. Physiol.}
{\bf 285}: H2437-H2445. \\
Kaas JH (2000)  Why is brain size so important: Design
problems and solutions as neocortex gets bigger or smaller. {\it Brain
Mind \/} {\bf 1}:  7-23.  \\
Kandel ER, Schwartz JH, Jessell TM (1991)
{\it Principles of Neural Science.}  Norwalk, Connecticut: 
Appleton and Lange, 3rd edition. \\
Karbowski J (2001) Optimal wiring principle and plateaus
in the degree of separation for cortical neurons. 
{\it Physical Review Letters} {\bf 86}: 3674-3677. \\
Karbowski J (2003) How does connectivity between cortical areas
depend on brain size? Implications for efficient computation.
{\it J. Comput. Neurosci.} {\bf 15}: 347-356. \\
Karbowski J (2007) Global and regional brain metabolic scaling and its
functional consequences. {\it BMC Biology} {\bf 5}: 18. \\
Kiyatkin EA (2007) Brain temperature fluctuations during physiological
and pathological conditions. {\it Eur. J. Appl. Physiol.} {\bf 101}:
3-17. \\
Koch C (1998) {\it Biophysics of computation.} 
Oxford: Oxford Univ. Press. \\
Koehler RC, Traystman RJ, Jones MD (1985) Regional blood flow and 
$O_{2}$ transport during hypoxic and CO hypoxia in neonatal and adult
sheep. {\it Am. J. Physiol.} {\bf 248}: H118-H124. \\
Landauer R (1961) Irreversibility and heat generation in the computing
process. {\it IBM J. Res. Dev.} {\bf 5}: 183-191. \\
Lennie P (2003) The cost of cortical computation. {\it Curr. Biol.}
{\bf 13}: 493-497. \\
Linde R, Schmalbruch IK, Paulson OB, Madsen PL (1999) The Kety-Schmidt
technique for repeated measurements of global cerebral blood flow
and metabolism in the conscious rat. {\it Acta Physiol. Scand.} 
{\bf 165}: 395-401. \\
Madsen PL et al (1991) Cerebral $O_{2}$ metabolism and cerebral blood
flow in humans during deep and rapid-eye-movement sleep. 
{\it J. Appl. Physiol.} {\bf 70}: 2597-2601. \\
Marcus ML, Heistad DD (1979) Effects of sympathetic nerves on cerebral
blood flow in awake dogs. {\it Am. J. Physiol.} {\bf 236}: H549-H553. \\
Markram H et al (1997) Physiology and anatomy of synaptic connections
between thick tufted pyramidal neurones in the developing rat
neocortex. {\it J. Physiol.} {\bf 500}: 409-440. \\
Markram H, Wang Y, Tsodyks M (1998) Differential signaling via the same
axon of neocortical pyramidal neurons. {\it Proc. Natl. Acad. Sci. USA}
{\bf 95}: 5323-5328. \\
McElligott JG, Melzack R (1967) Localized thermal changes evoked in
the brain by visual and auditory stimulation. {\it Exp. Neurol.}
{\bf 17}: 293-312. \\
Murre JMJ, Sturdy DPF (1995) The connectivity of the brain:
Multilevel quantitative analysis. {\it Biol. Cybern.} {\bf 73}:
529-545. \\
Nakao M, Gadsby DC (1989) [Na] and [K] dependence of the Na/K pump
current-voltage relationship in guinea pig ventricular myocytes.
{\it J. Gen. Physiol.} {\bf 94}: 539-565. \\
Nelson DA, Nunneley SA (1998) Brain temperature and limits on transcranial
cooling in humans: quantitative modeling results. {\it Eur. J. Appl.
Physiol.} {\bf 78}: 353-359. \\
Nybo L, Secher NH, Nielsen B (2002) Inadequate heat release from the
human brain during prolonged exercise with hyperthermia. {\it J. Physiol.}
{\bf 545}: 697-704. \\
Prothero (1997) Cortical scaling in mammals: A repeating units
model. {\it J. Brain Res.} {\bf 38}: 195-207. \\
Purves D (1988) {\it Body and Brain.} 
Cambridge, Massachusetts: Harvard Univ. Press. \\
Raichle ME (2003) Functional brain imaging and human brain function.
{\it J. Neurosci.} {\bf 23}: 3959-3962. \\
Rolfe DFS, Brown GC (1997) Cellular energy
utilization and molecular origin of standard metabolic rate in mammals.
{\it Physiol. Rev.} {\bf 77}: 731-758. \\
Schmidt-Nielsen K (1984) {\it Scaling: Why is Animal Size
so Important?} Cambridge: Cambridge Univ. Press. \\
Siesjo B (1978) {\it Brain Energy Metabolism}. New York: Wiley. \\
Stephan H, Baron G, Frahm HD (1981) New and revised data on volumes of
brain structures in insectivores and primates. {\it Folia Primatol.}
{\bf 35}: 1-29. \\
Stowe K (1984) {\it Introduction to Statistical Mechanics and Thermodynamics}.
New York: Wiley. \\
Striedter GF (2005) {\it Principles of Brain Evolution}. 
Sunderland, MA: Sinauer Assoc. \\
Sukstanskii AL, Yablonskiy DA (2006) Theoretical model of temperature
regulation in the brain during changes in functional activity.
{\it Proc. Natl. Acad. Sci. USA} {\bf 103}: 12144-12149. \\
Traub R, Miles R (1991) {\it Neuronal Networks of the Hippocampus}.
Cambridge, UK: Cambridge Univ. Press. \\
Waschke K et al (1993) Local cerebral blood flow and glucose utilization
after blood exchange with a hemoglobin-based $O_{2}$ carrier in conscious
rats. {\it Am. J. Physiol.} {\bf 265}: H1243-H1248. \\
van Leeuwen GMJ et al (2000) Numerical modeling of temperature distribution
within the natal head. {\it Pediatr. Res.} {\bf 48}: 351-356. \\
Volgushev M, et al (2004) Probability of transmitter release at neocortical
synapses at different temperatures. {\it J. Neurophysiol.} {\bf 92}: 212-220. \\
Yablonskiy DA, Ackerman JJH, Raichle ME (2000) Coupling between changes in
human brain temperature and oxidative metabolism during prolonged visual
stimulation. {\it Proc. Natl. Acad. Sci. USA} {\bf 97}: 7603-7608. \\
Yoshinura Y, Kimura F, Tsumoto T (1999) Estimation of single channel conductance
underlying synaptic transmission between pyramidal cells in the visual cortex.
{\it Neuroscience} {\bf 88}: 347-352. \\
Zhang K, Sejnowski TJ (2000) A universal scaling law between gray matter
and white matter of cerebral cortex. {\it Proc. Natl. Acad. Sci. USA}
{\bf 97}: 5621-5626.



\newpage

{\bf \large Figure Captions}

Fig. 1\\
Heat transfer in the brain.
Brain is represented as half of the ball with three concentric 
layers representing (from the brain outside): cerebrospinal fluid, skull, 
and scalp. The heat generated in the brain is removed by cerebral 
blood flow (small circles), conduction through the brain and three layers 
(dashed arrows), and scalp convection/conduction and radiation 
(dashed circular line).

\vspace{0.3cm}

Fig. 2\\
Sodium influx during an action potential. 
(A) Temporal dependence of Na$^{+}$ (solid line) and K$^{+}$ (dashed
line) conductances and voltage (dashed-dotted line) during the action
potential. (B) The correction
$\delta C$ as a function of intracellular sodium concentration. 
(C) Dependence of sodium influx $\Delta[\mbox{Na}]_{o}$ on the effective
fiber diameter $d$ coming from a numerical integration of Eq. (6). 
The least square fit (solid line) yields the relationship: 
$\Delta[\mbox{Na}]_{o}= 2.72\cdot 10^{-5} d^{-0.974}$ mM,
where $d$ is in cm. This numerical fit practically confirms the 
theoretical dependence $\Delta[\mbox{Na}]_{o}\sim d^{-1}$ in Eq. (12).

\vspace{0.3cm}

Fig. 3\\
Dependence of the intracellular sodium concentration on time
for a neuron firing repeatedly. (A) Results for $d= 0.45$ $\mu$m, i.e.,
the harmonic mean of the mouse axon diameter $d_{a}=0.3$ $\mu$m and
dendrite diameter $d_{d}=0.9$ $\mu$m (see Sec. 2.2).
(B) Results for $d= 0.11$ $\mu$m, i.e., for the smallest molecularly
possible axons with $d_{a}= 0.06$ $\mu$m. Note that the range of variability
in (B) is much larger, which reflexs higher amplitudes of Na$^{+}$
influx and its faster relaxation for thin fibers.
Panels (A) and (B) come from simulations of Eq. (6).
(C) Simulation results for $d= 0.45$ $\mu$m of Eq. (13) (solid lines) 
and its approximation Eq. (14) (dashed and dashed-dotted lines).
For all plots in (A)-(C) $k= 3$.

\vspace{0.3cm}

Fig. 4\\
Intracellular sodium concentration as a function of firing rate. 
The point (diamonds) coming from a direct numerical integration 
of Eq. (6) are approximated well by the theoretical formula (15) 
(solid line). The maximal discrepancy for high frequency is about 
$15-20$ $\%$. The average [Na]$_{av}$ does not depend much on fiber 
diameter (panels A and C) nor on the Hill coefficient $k$ (panels A and B).

\vspace{0.3cm}

Fig. 5\\
ATP and glucose utilization rates, and scaling of firing rate
with brain size.
(A) ATP rate increases non-linearly with frequency $f$ and saturates for
high values of $f$. (B) CMR$_{glu}$ also increases non-linearly with
firing rate, similar to the dependence of ATP on $f$.
(C) Scaling of the estimated firing rates in mammals with brain size.
The least square fit to the data points yields 
$f_{est}= 4.79 U_{g}^{-0.15}$ Hz  ($R^{2}=0.89$, $p= 0.0015$). 
Estimates of frequency are based on the empirical data of CMR$_{glu}$,
which are (in $\mu$mol/cm$^{3}$min): 1.07 for mouse, 0.90 for rat, 0.83
for rabbit, 0.81 for cat, 0.47 for rhesus monkey, 0.46 for baboon, 0.34
for human. These data were taken from the data gathered in the 
supplementary information of Karbowski (2007). 
For all panels: $d= 0.45$ $\mu$m, $k= 3$.

\vspace{0.3cm}

Fig. 6\\
Dependence of the sodium pump power $P_{ATP}$ on frequency and the effective
fiber diameter. (A) Biphasic dependence of $P_{ATP}$ (per neuron per surface 
area) on frequency. Analytical formula (19) (solid line) leads to similar 
results as a direct numerical integration of Eq. (6) (diamonds). 
Results are for $d= 0.45$ $\mu$m.
(B) $P_{ATP}$ depends inversely on fiber diameter $d$. For 
extremely thin fibers $P_{ATP}$ tends to diverge. Results are for 
$U_{g}= 680$ cm$^{3}$ and $f= 1$ Hz, corresponding to human brain.
In both panels A and B, the Hill coefficient $k= 3$.

\vspace{0.3cm}

Fig. 7\\
Log-Log plot of the empirical dependence of the cerebral blood flow 
rate $CBF$ on brain volume $U_{br}$ for several mammals. The allometric 
dependence has the following form: $CBF= 0.018 U_{br}^{-0.10}$ 1/sec,
where $U_{br}$ is expressed in cm$^{3}$
($R^{2}= 0.976$, $p= 0.0002$).
The $CBF$ empirical data are as follows: 
$20.2\cdot 10^{-3}$ sec$^{-1}$ for mouse (Frietsch et al, 2007); 
$17.7\cdot 10^{-3}$ sec$^{-1}$ for rat 
(Waschke et al, 1993; Linde et al, 1999); 
$14.3\cdot 10^{-3}$ sec$^{-1}$ for rabbit (Busija, 1984);
$12.7\cdot 10^{-3}$ sec$^{-1}$ for dog (Marcus and Heistad, 1979); 
$10.7\cdot 10^{-3}$ sec$^{-1}$ for sheep (Koehler et al, 1985); 
$9.2\cdot 10^{-3}$ sec$^{-1}$ for human (Madsen et al, 1991).
Brain volumes were taken from Stephan et al (1981) and Karbowski (2007).

\vspace{0.3cm}

Fig. 8\\
Brain temperature as a function of brain size, fiber diameter, and
firing rate. (A) The spatial distribution of brain temperature $T(r)$.
For very small brains $T(r)$ is smaller than the arterior 
blood temperature $T_{bl}$ (dotted line).
(B), (C) $T(0)$ strongly increases for very thin fibers 
($U_{g}= 680$ cm$^{3}$ corresponding to human brain for panel B and C; 
and $f= 1.7$ Hz for panel B). 
(D) Biphasic dependence of $T(0)$ on firing rate ($d= 0.45$ $\mu$m, $k= 3$).
Solid line corresponds to human and dashed line to mouse.

\vspace{0.3cm}

Fig. 9\\
Lower bound on the effective fiber diameter $d$ as a function of brain size and
firing rate. 
(A) $d_{min}$ depends very weakly on gray matter volume.
(B) The lower bound on $d_{min}$ depends biphasically on firing rate
($U_{g}= 680$ cm$^{3}$). Weak synaptic depression ($\gamma= 0.1$; dashed
line) leads to a similar dependence as moderate synaptic depression
($\gamma= 0.5$; solid line).
The lower thermal bound on axon diameter $d_{a}$ is approximately two times
smaller than $d_{min}$, and therefore it is maximally $0.026$ $\mu$m.

\newpage

\begin{table}
\begin{center}
\caption{Parameters used in the article.}
\begin{tabular}{|l l l l|}
\hline
\hline
Parameter &  Value   &  Units  &  Reference/Comment         \\
\hline
\hline

$g_{Na,o}$ & $2.9\cdot 10^{-7}$  &  ($\Omega$cm$^{2}$)$^{-1}$
&  at $-67$ mV Traub-Miles model (1991)   \\ 
$g_{K,o}$  & $1.2\cdot 10^{-6}$  &  ($\Omega$cm$^{2}$)$^{-1}$
&  at $-67$ mV Traub-Miles model (1991)   \\
$\overline{g}_{Na}$ & $0.1$  &  ($\Omega$cm$^{2}$)$^{-1}$
& Traub-Miles model (1991)   \\ 
$\overline{g}_{K}$  & $0.085$  &  ($\Omega$cm$^{2}$)$^{-1}$
&  Traub-Miles model (1991)   \\
$g_{L}$  & $10^{-4}$  &  ($\Omega$cm$^{2}$)$^{-1}$
&  Traub-Miles model (1991)  \\
$g_{s}$ & $0.3\cdot 10^{-9}$  &    $\Omega^{-1}$ 
& Yoshimura et al (1999) \\
$q_{0}$ & 0.17     &  unitless & Volgushev et al (2004) \\
 &  &   & Markram et al (1998) \\
 &  &   & Yoshimura et al (1999) \\

$V_{s}$ & 0.000 & V 
&  Koch (1998) \\ 
$C$ & $10^{-6}$ & F/cm$^{2}$ 
&  generally accepted \\

$\tau_{s}$ & $2.2\cdot 10^{-3}$ & sec 
&  Koch (1998) \\ 
$\tau_{d}$ & $0.5$ & sec 
&  Dayan and Abbott (2001)  \\ 
$\gamma$ & $0.5$ & unitless 
&  Dayan and Abbott (2001) \\

$A$ & $2\cdot 10^{-6}$ &  A/cm$^{2}$ &   Abercrombie and De Weer (1978) \\
 &   &   &  Nakao and Gadsby (1989) \\

$\rho_{s}$ &  $5\cdot 10^{11}$  & cm$^{-3}$ 
& Braitenberg and Schuz (1998) \\
 &   &   & Craig (1967) \\
 &   &   & DeFelipe et al (2002) \\
$\phi$ & 1/3 & unitless & Braitenberg and Schuz (1998) \\

$F$ &  96485.3  &  C/mol  
&  Faraday constant \\

$\kappa$ &  $5\cdot 10^{-3}$  & W/(cm K) 
&  van Leeuwen et al (2000) \\ 
$\sigma_{SB}$ &  $5.67\cdot 10^{-12}$  &  W/(cm$^{2}$K$^{4}$) 
&  Stowe (1984) \\ 
$\eta$ &  $1.2\cdot 10^{-3}$  &  W/(cm K)
& Nelson and Nunneley (1998) \\
 &   &   &  van Leeuwen et al (2000) \\ 
$c_{bl}$ &  $3.8\cdot 10^{3}$  & J/(kg K) 
&  generally accepted \\
$\rho_{bl}$ &  $1.06\cdot 10^{-3}$  & kg/cm$^{3}$ 
&  generally accepted \\
$T_{bl}$ &  309.8  & K 
& cerebral blood (body) temperature  \\
$T_{o}$ &  293.2  & K 
&  room temperature (20 $^{o}$C) \\

\hline

\hline
\end{tabular}

\end{center}

\end{table}

\newpage

\begin{table}
\begin{center}
\caption{Estimated neural activities and thermal brain properties in
several mammals.}
\begin{tabular}{|l l l l l l l l l |}
\hline
\hline
Species &  $f$ (Hz) & $P_{ATP}$ (W)  &  $T(0)$ ($^{o}$C)  &  $T_{sc}$ ($^{o}$C)
&  $\dot{Q}_{bl}$  (W)  &  $\dot{Q}_{c}$  (W)  &   $\dot{Q}_{cv}$  (W)  &   
$\dot{Q}_{r}$  (W)    \\
\hline
\hline

Mouse  &  6.18  & 0.003 &  36.57 &  35.5  &  $-0.024$ & 0.027 & 0.017 &  0.009  \\ 
Rat  &    5.03  & 0.008 &  36.69 &  35.4  &  $-0.058$ & 0.066 & 0.043 &  0.022  \\ 
Rabbit &  4.59  & 0.054 &  36.81 &  35.3  &  $-0.199$ & 0.25  & 0.167 &  0.086  \\
Cat &     4.47  & 0.27  &  36.85 &  35.2  &  $-0.51$  & 0.78  & 0.51  &  0.26   \\
Macaque & 2.38  & 0.53  &  36.76 &  35.0  &  $-1.25$  & 1.78  & 1.18  &  0.60   \\
Baboon &  2.33  & 0.84  &  36.76 &  35.0  &  $-1.65$  & 2.49  & 1.64  &  0.84   \\
Human &   1.68  & 5.41  &  36.73 &  34.7  &  $-6.02$  & 11.43 & 7.55  &  3.88   \\

\hline

\hline
\end{tabular}

\end{center}

\noindent In estimations it was assumed that blood temperature $T_{bl}= 36.6$
$^{o}C$, and it is brain size independent. The following values of gray
matter volume $U_{g}$ were used: 0.11 cm$^{3}$ for mouse; 0.42 cm$^{3}$ for
rat; 3.0 cm$^{3}$ for rabbit; 15.2 cm$^{3}$ for cat; 50.0 cm$^{3}$ for macaque
monkey; 80.0 cm$^{3}$ for baboon; and 680.0 cm$^{3}$ for human.

\end{table}

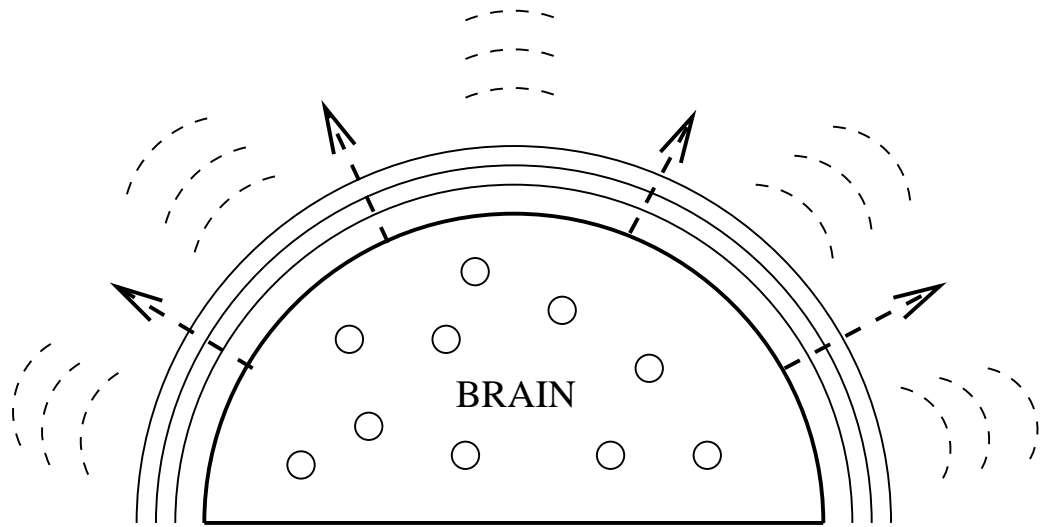
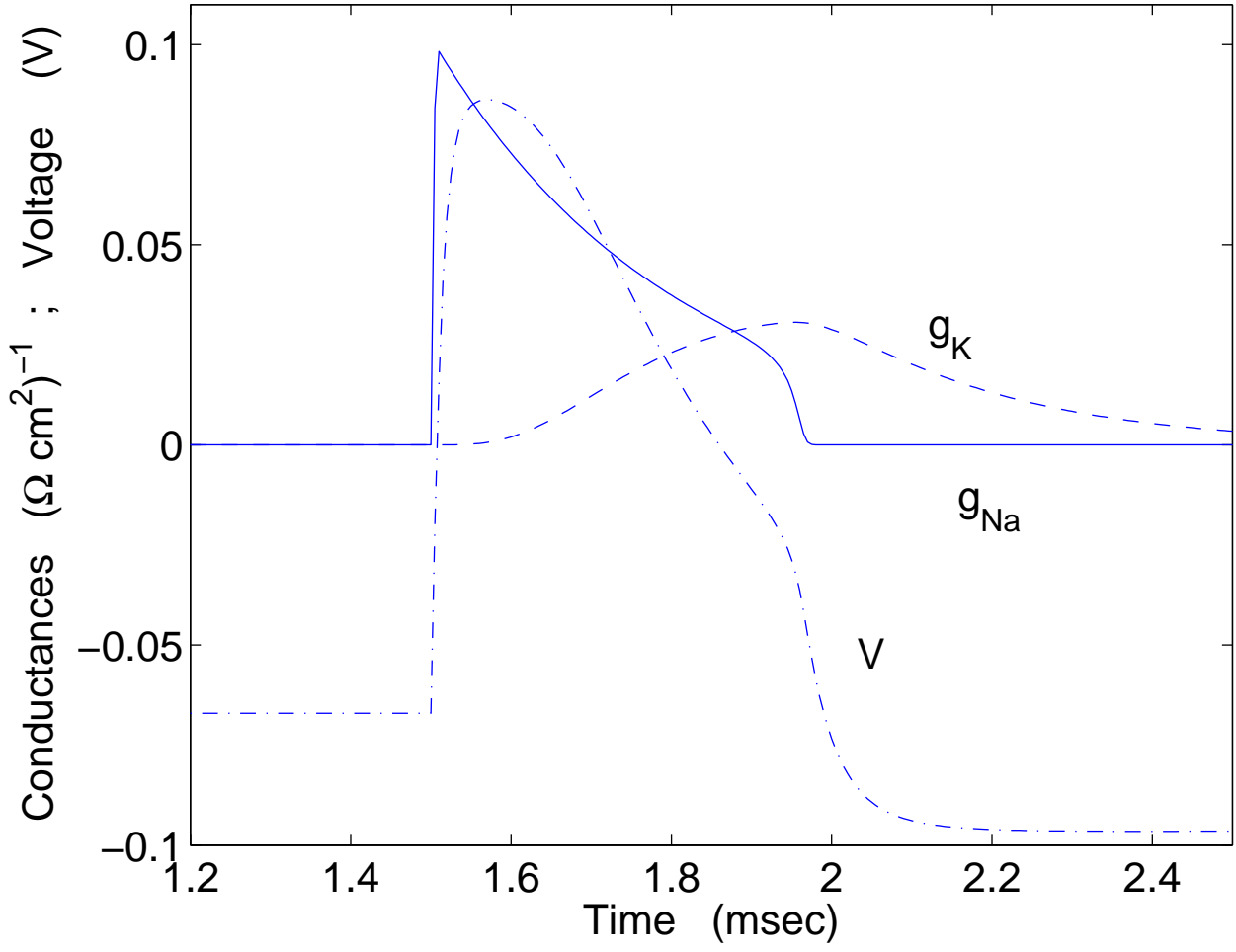
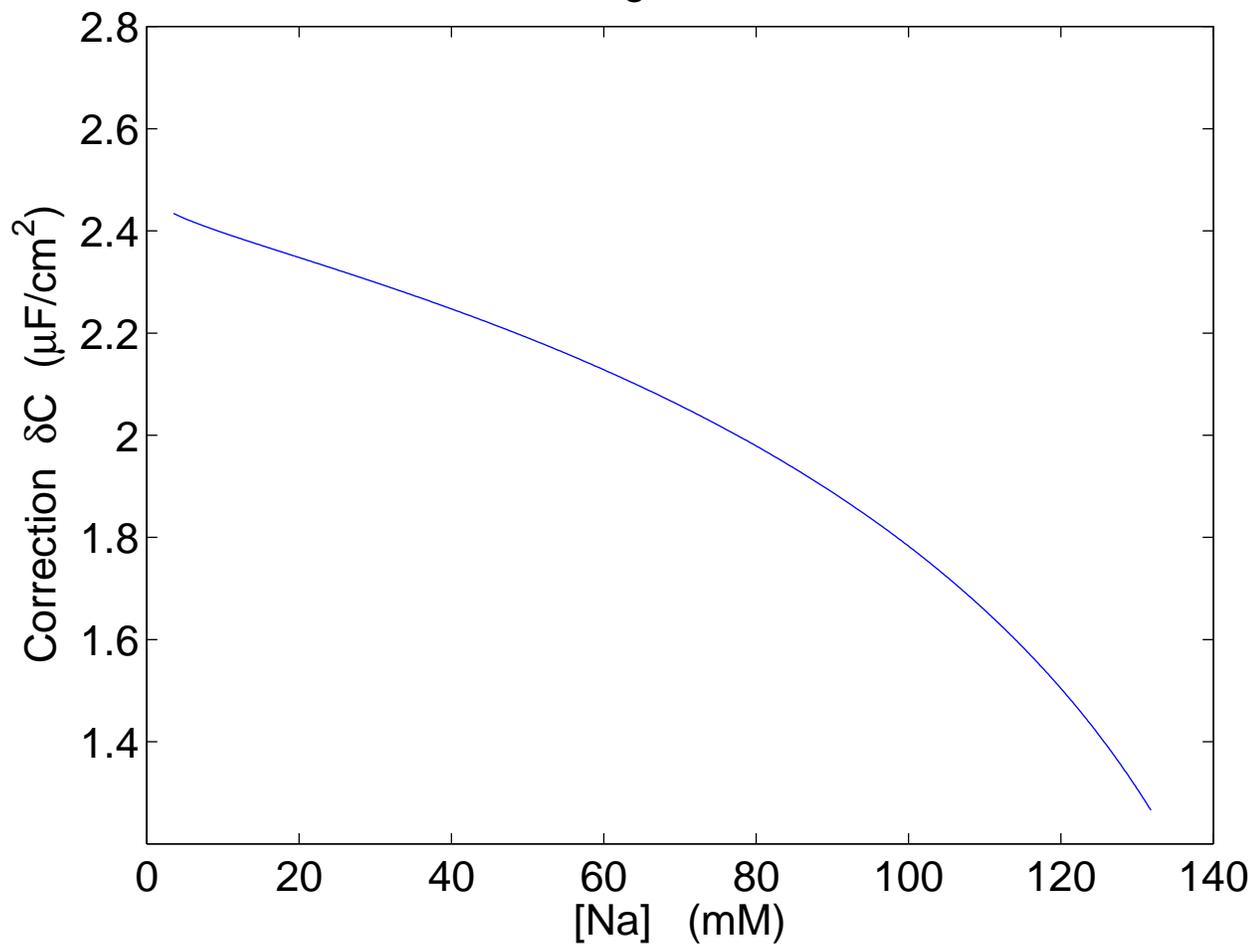
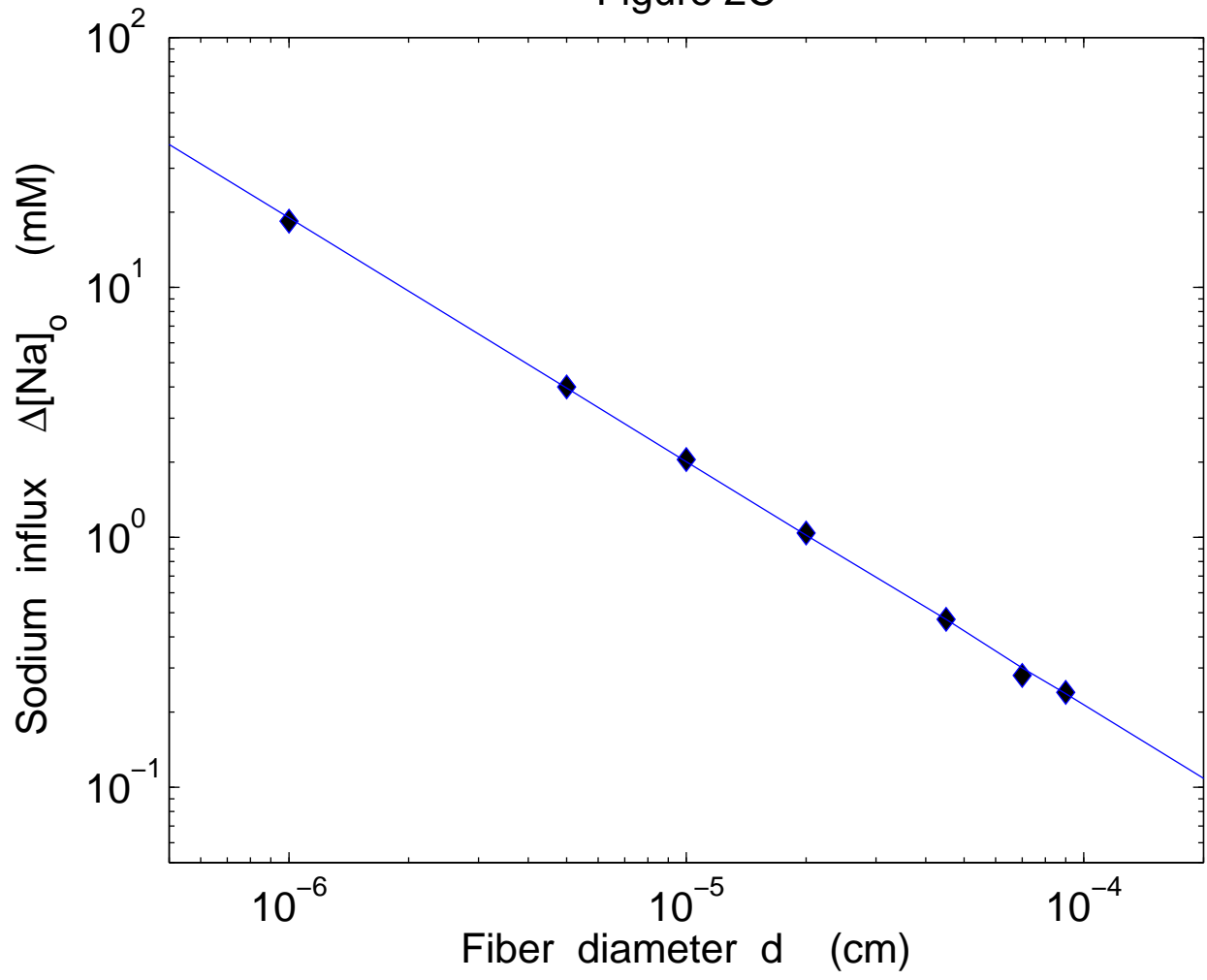
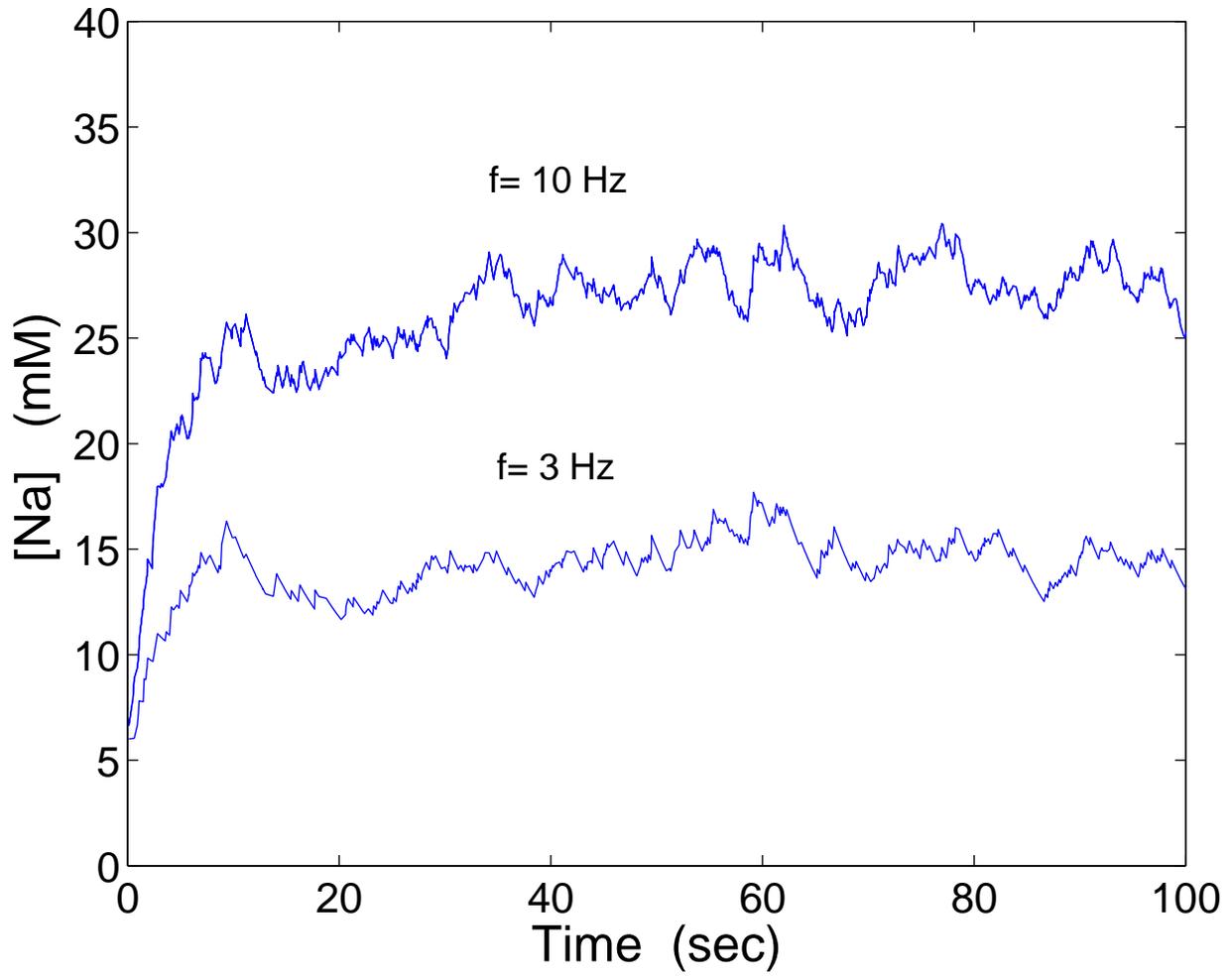
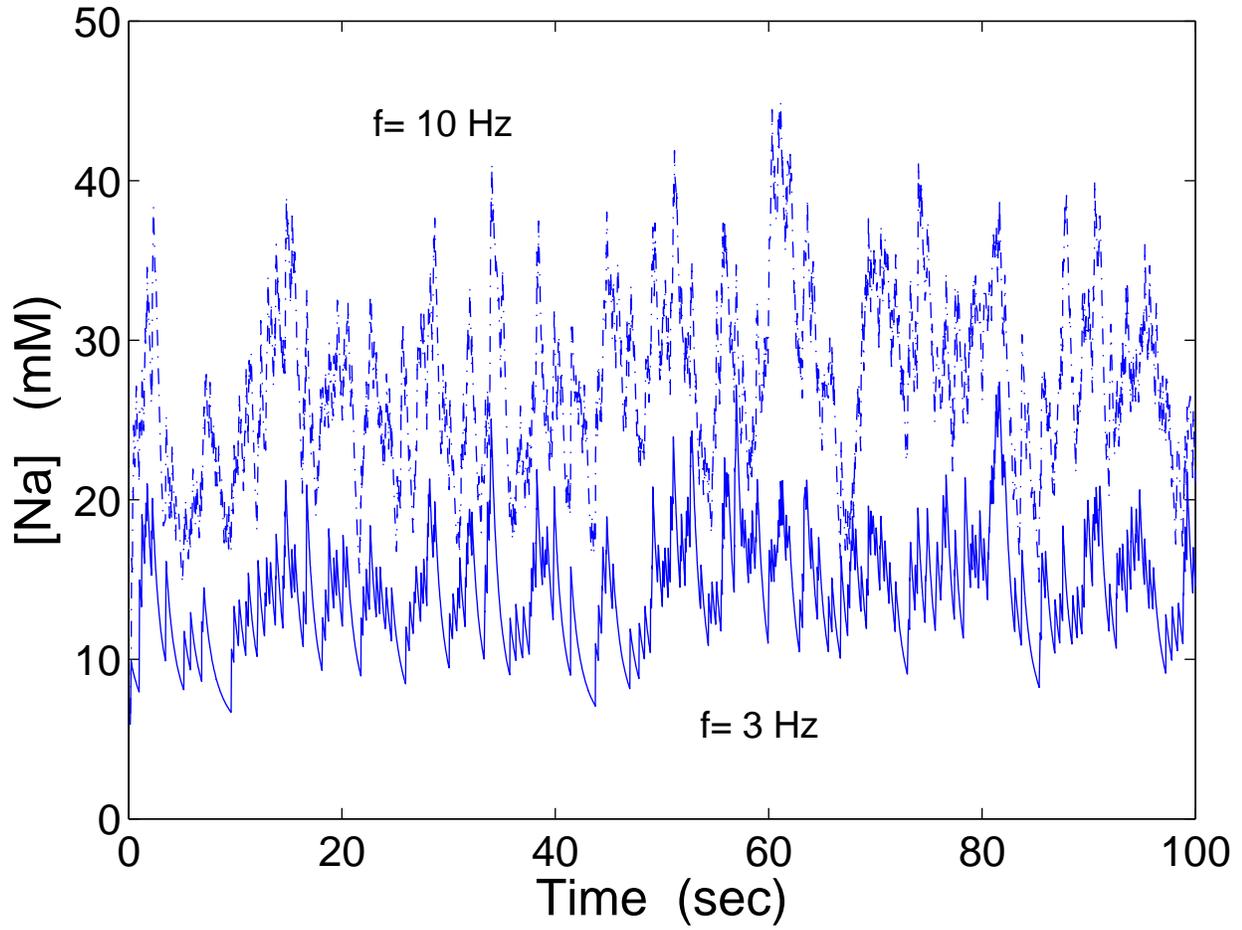
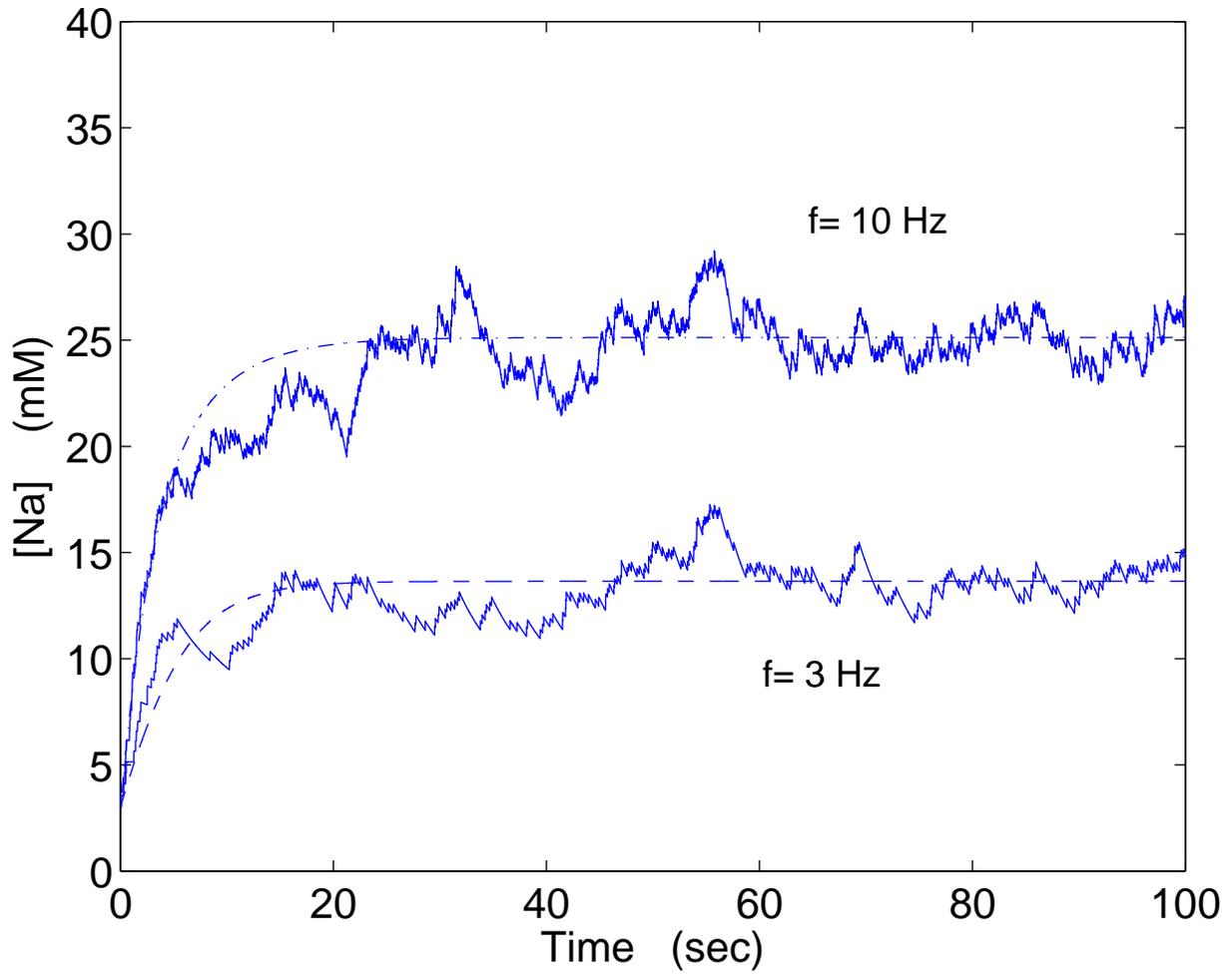
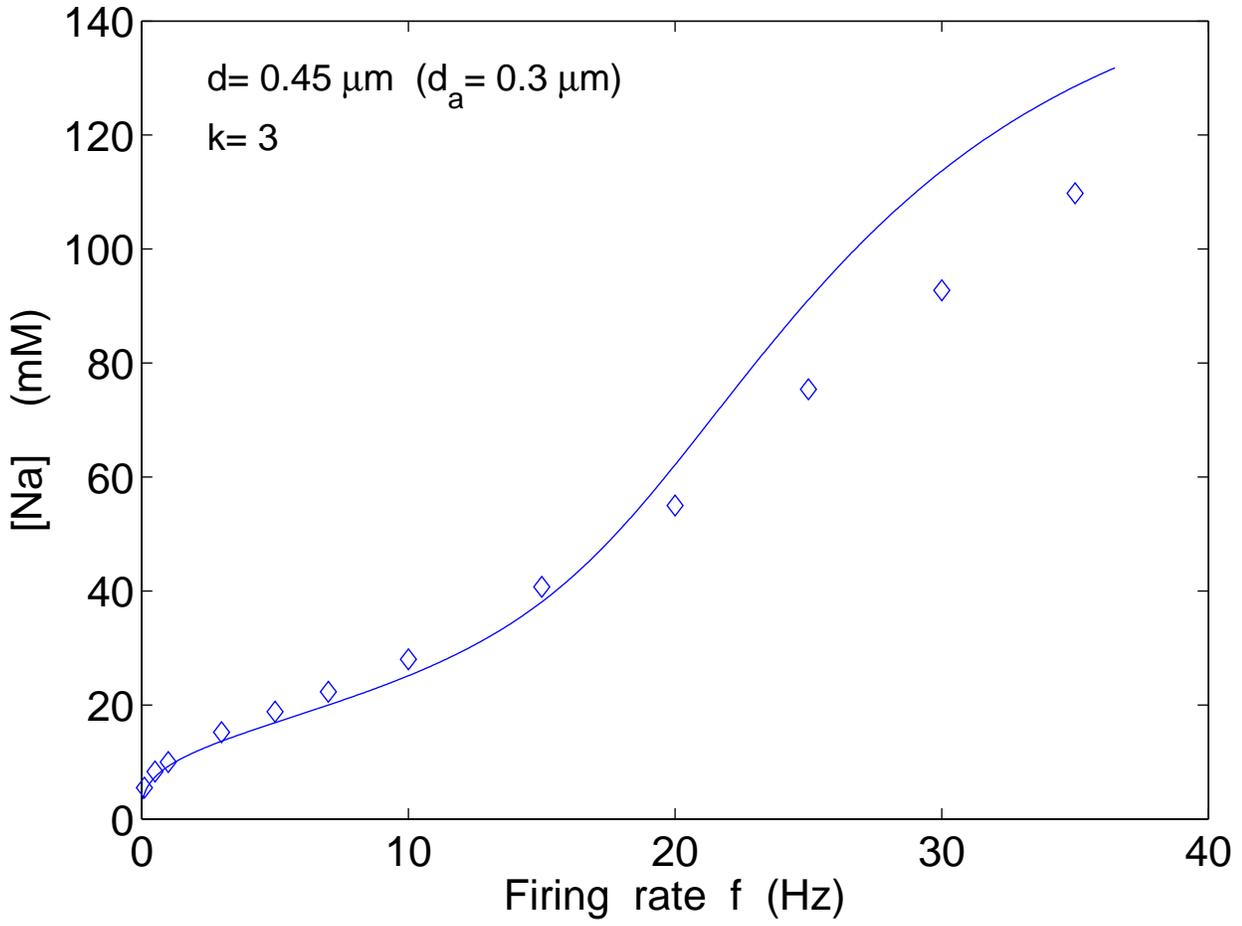
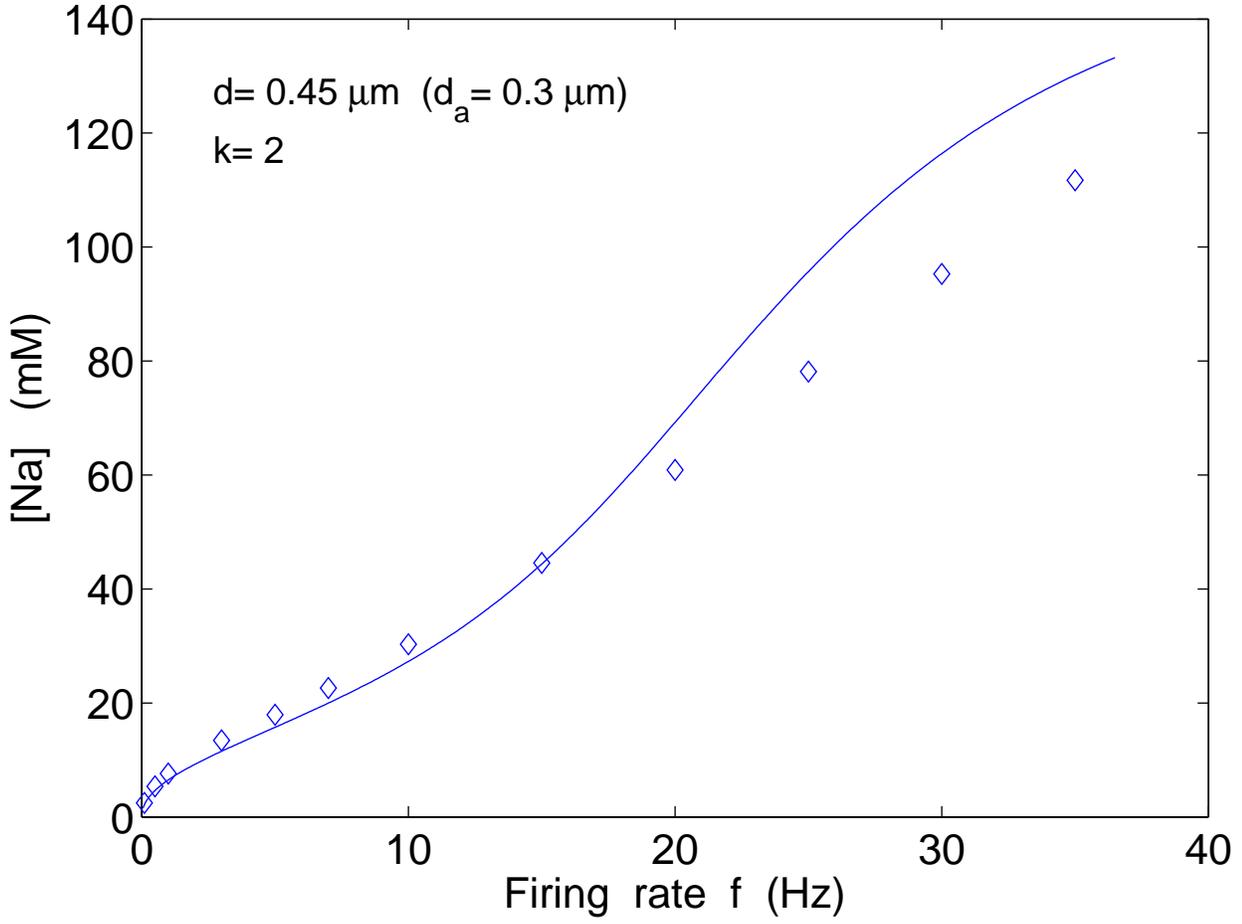
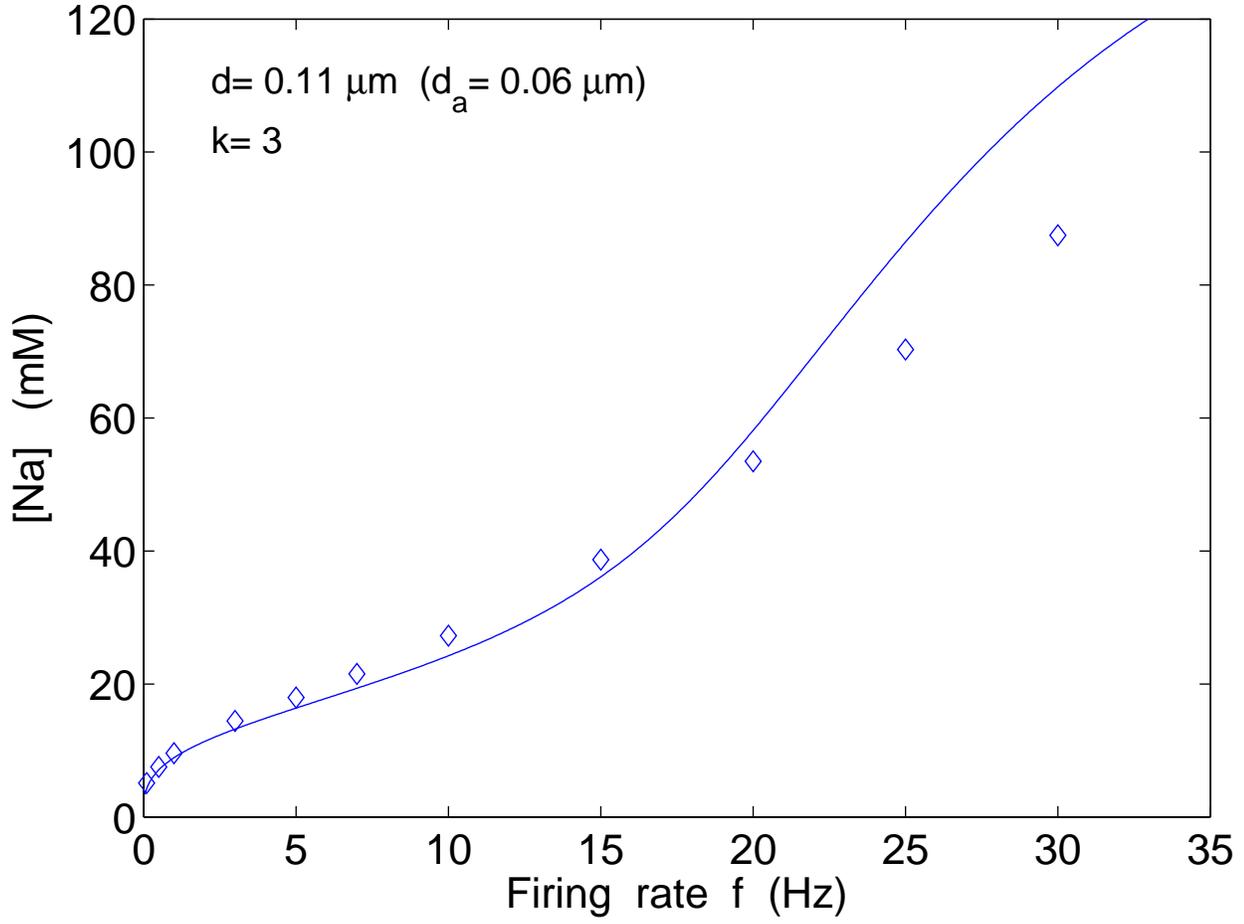
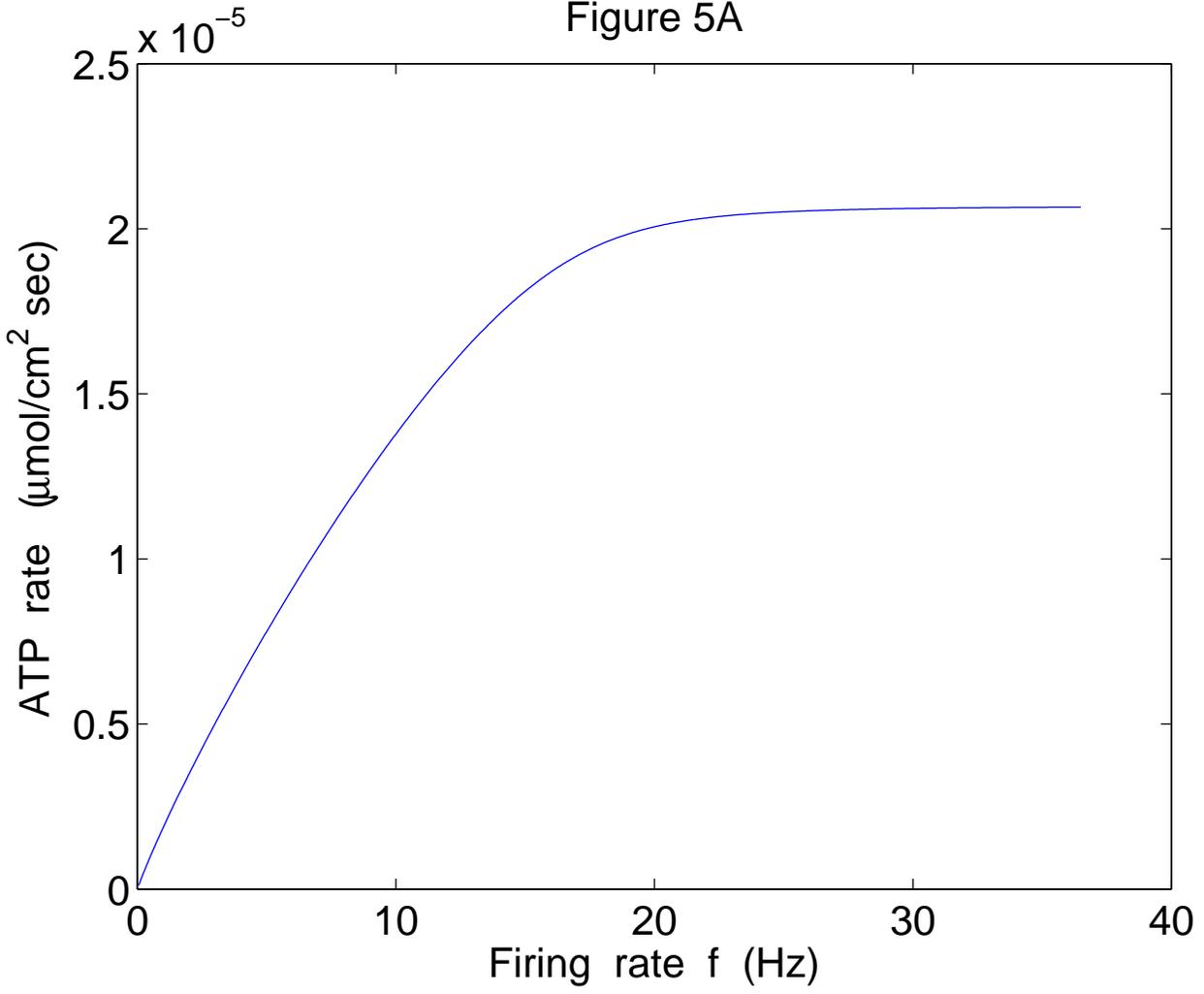
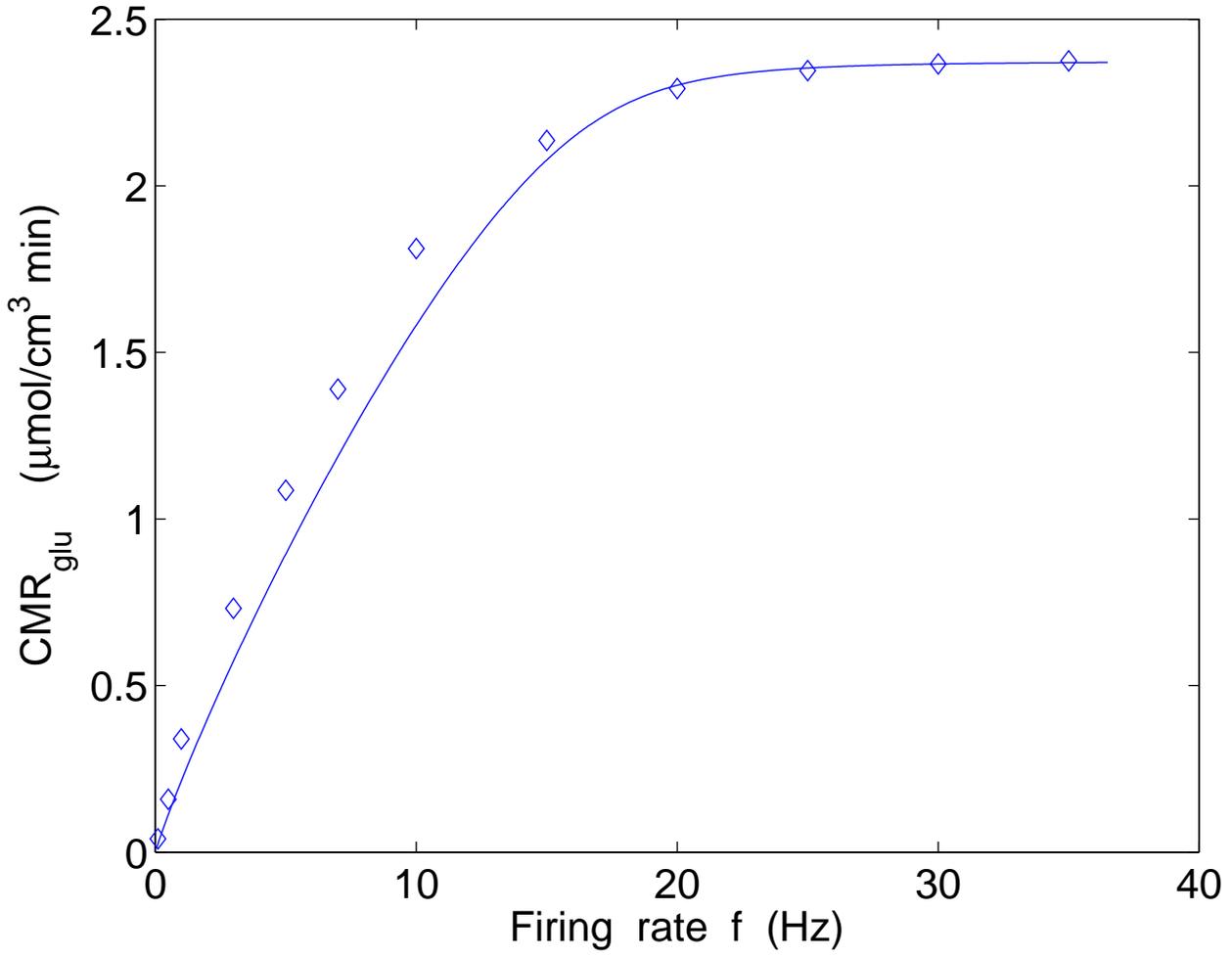
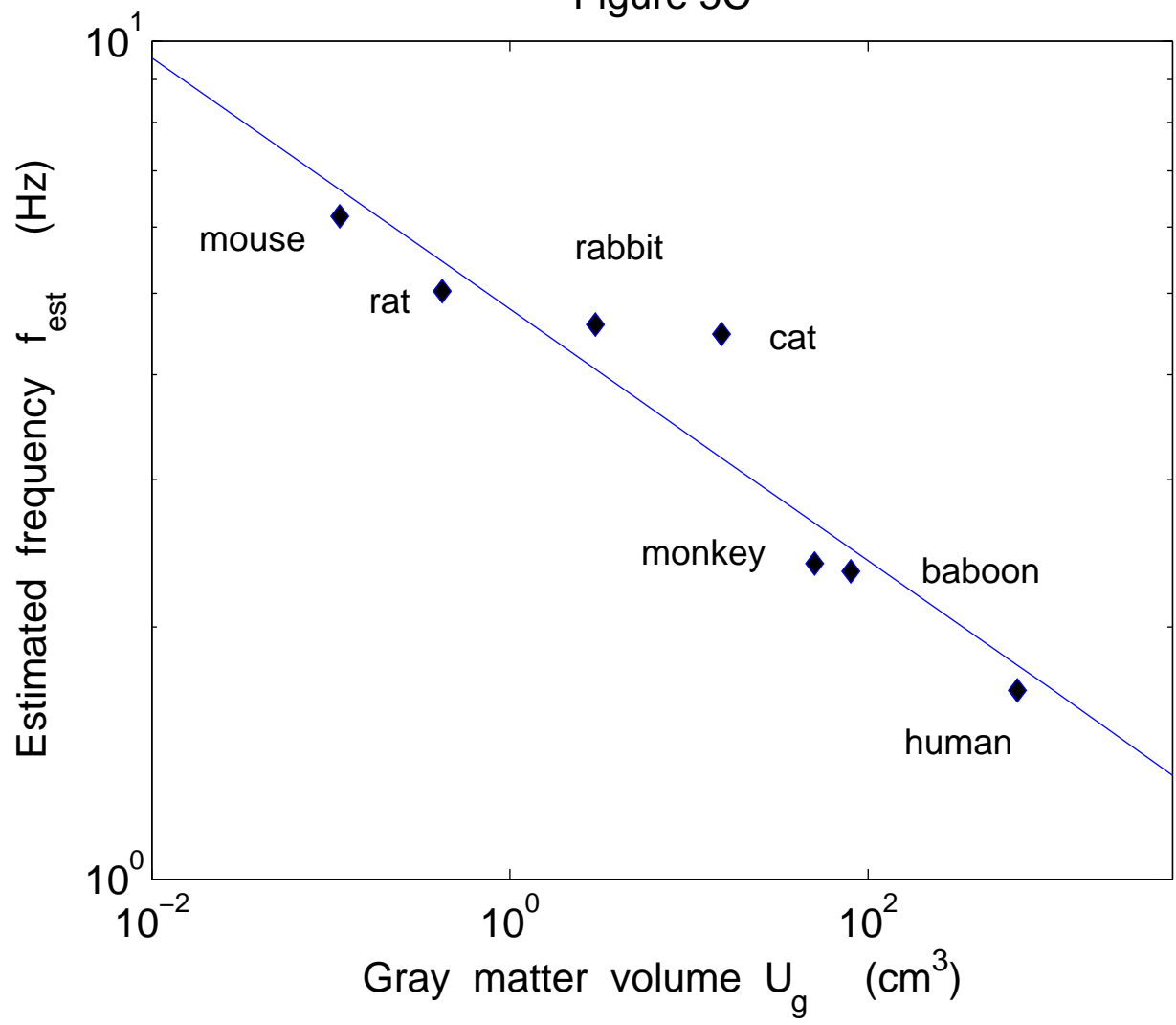
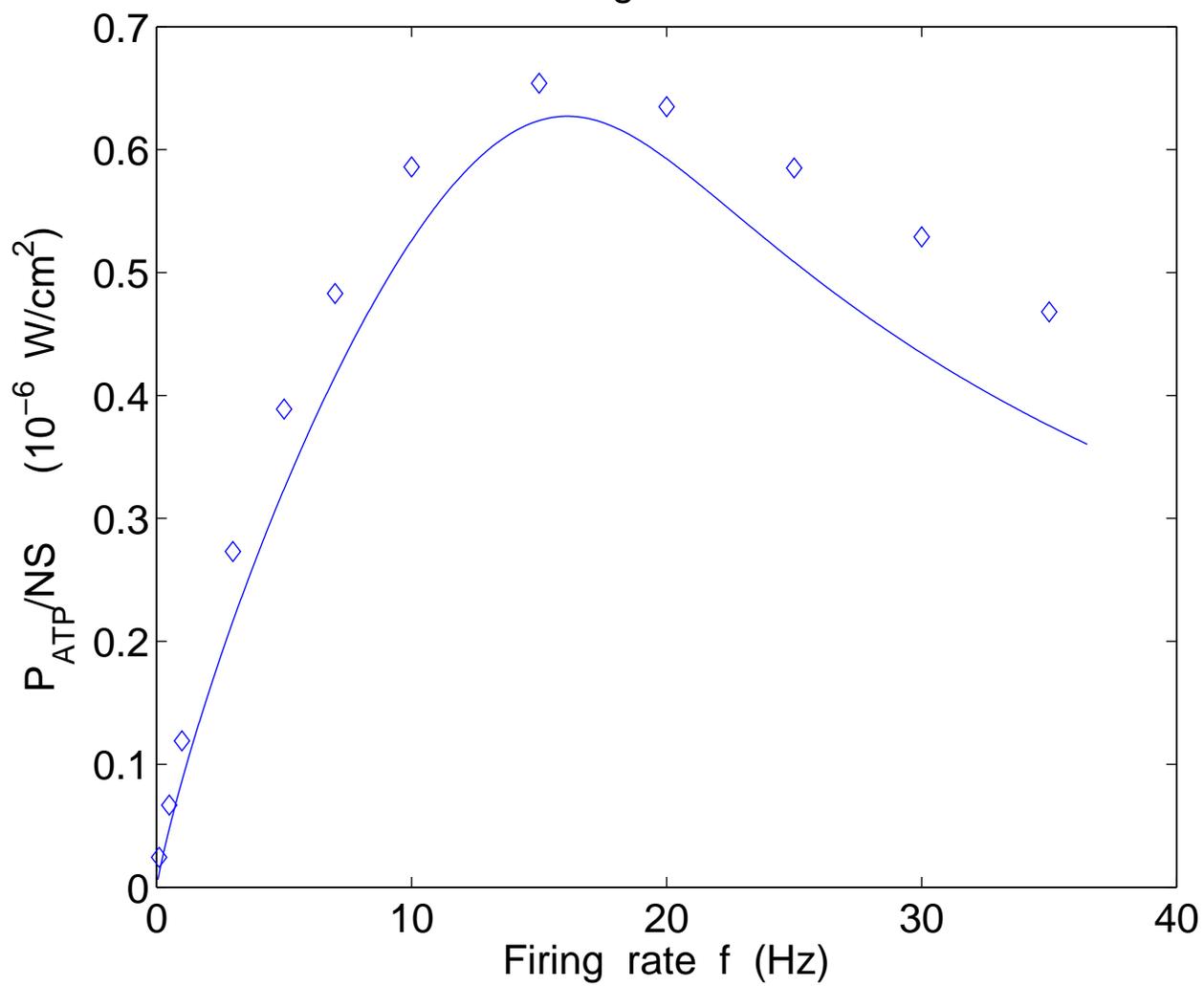
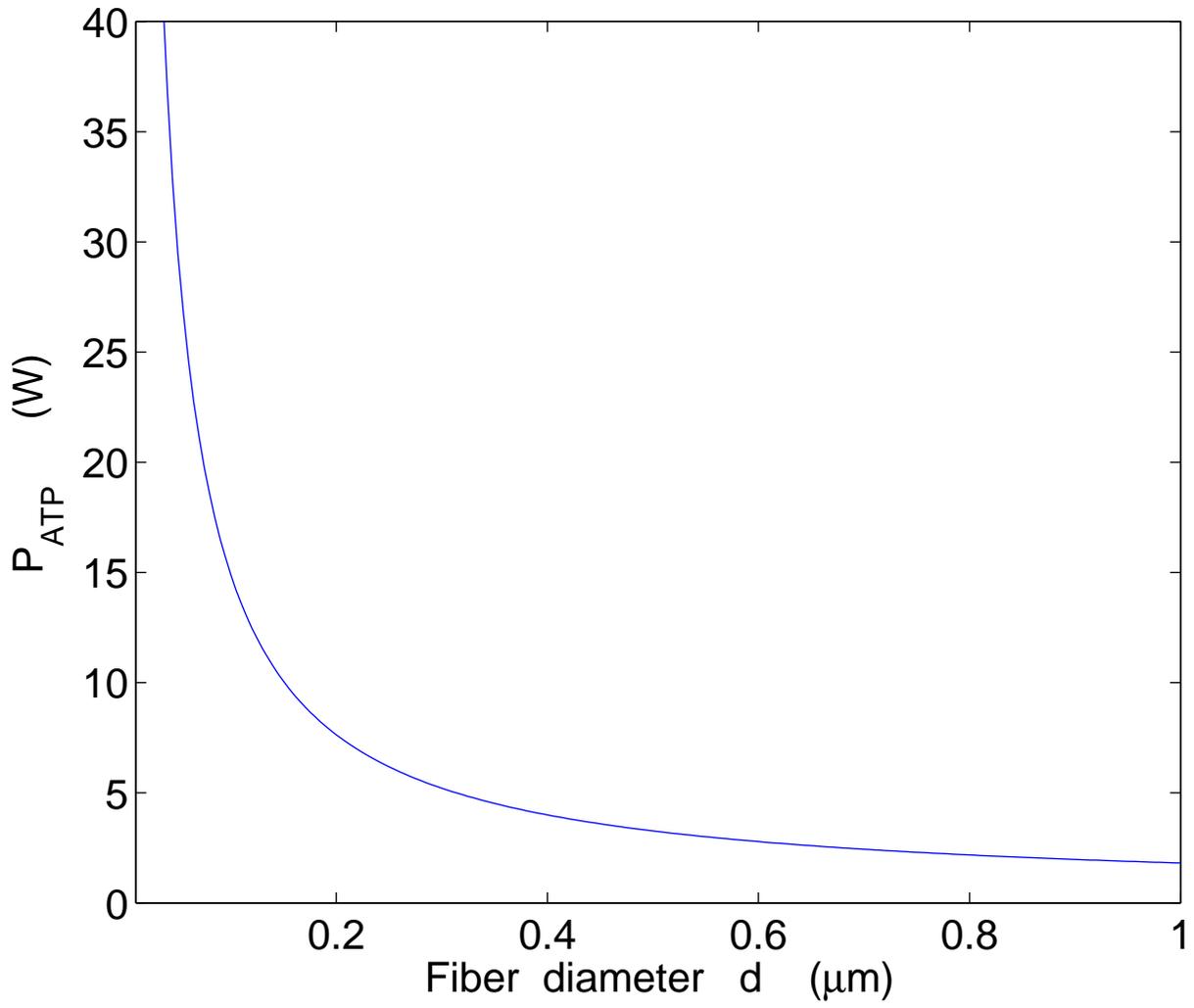
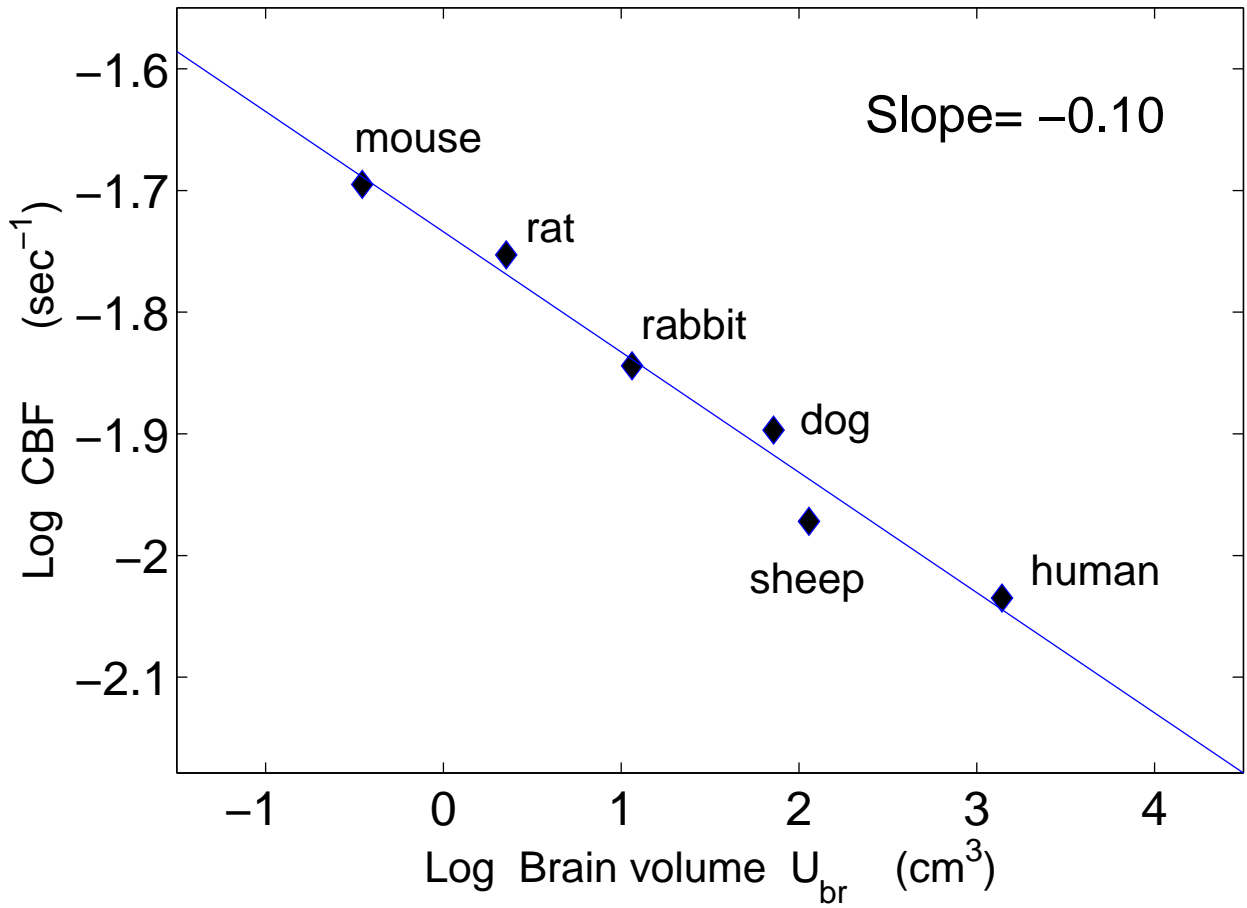
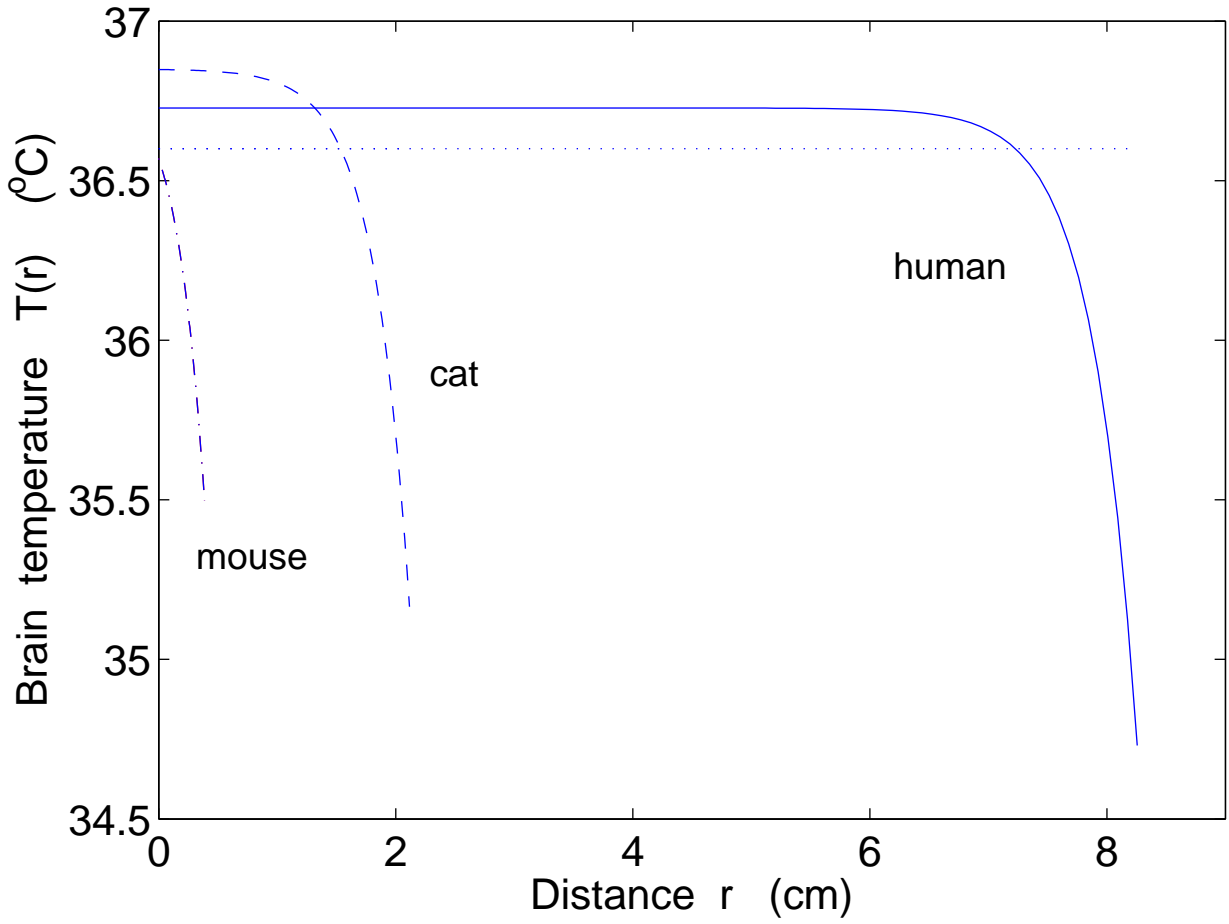
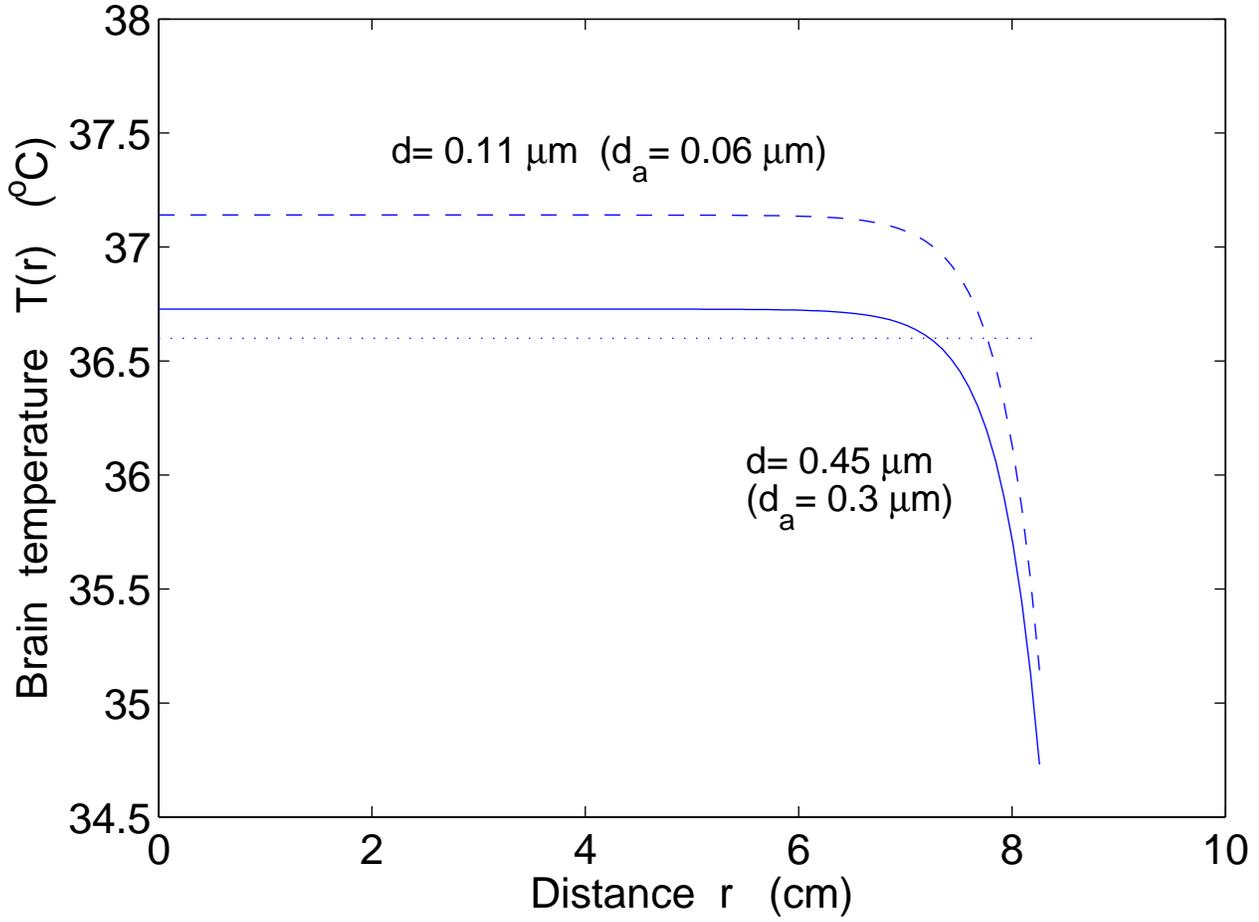
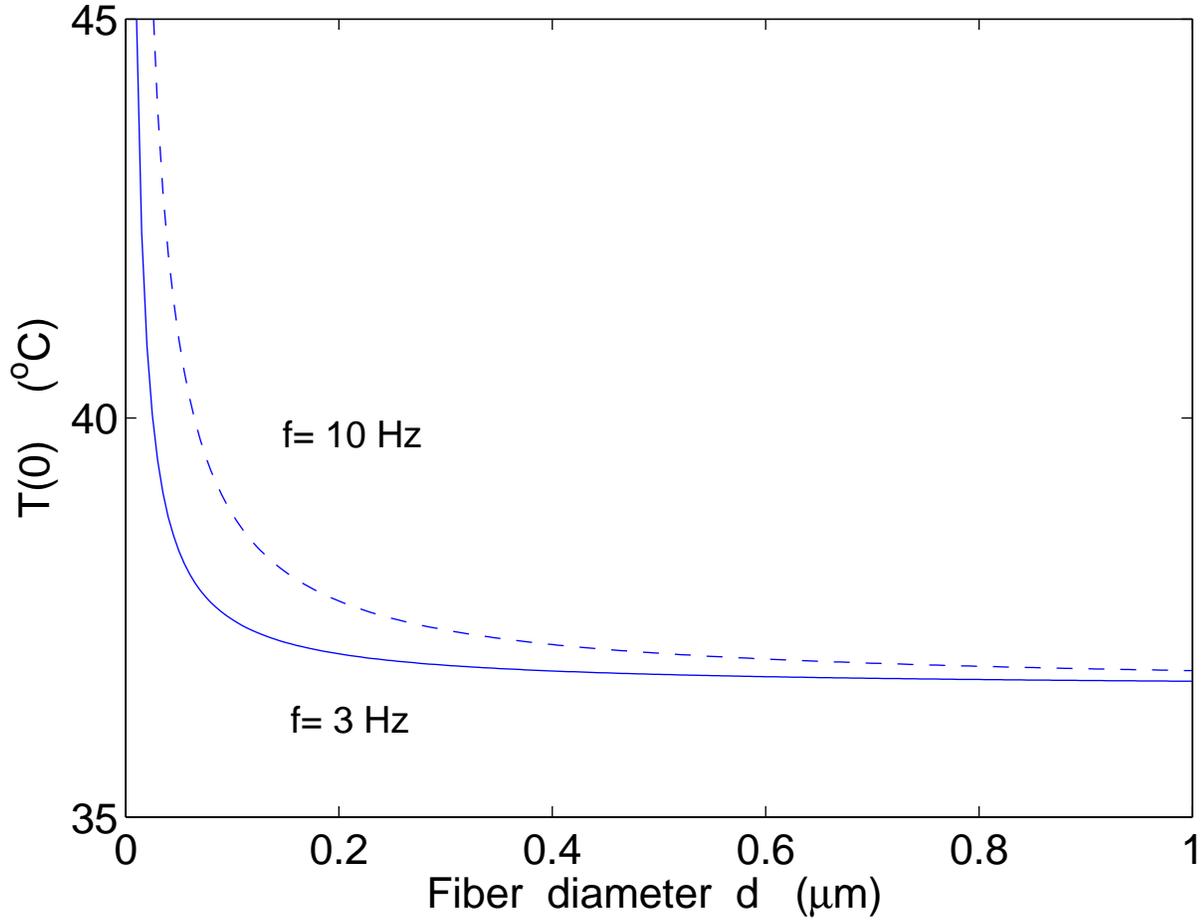
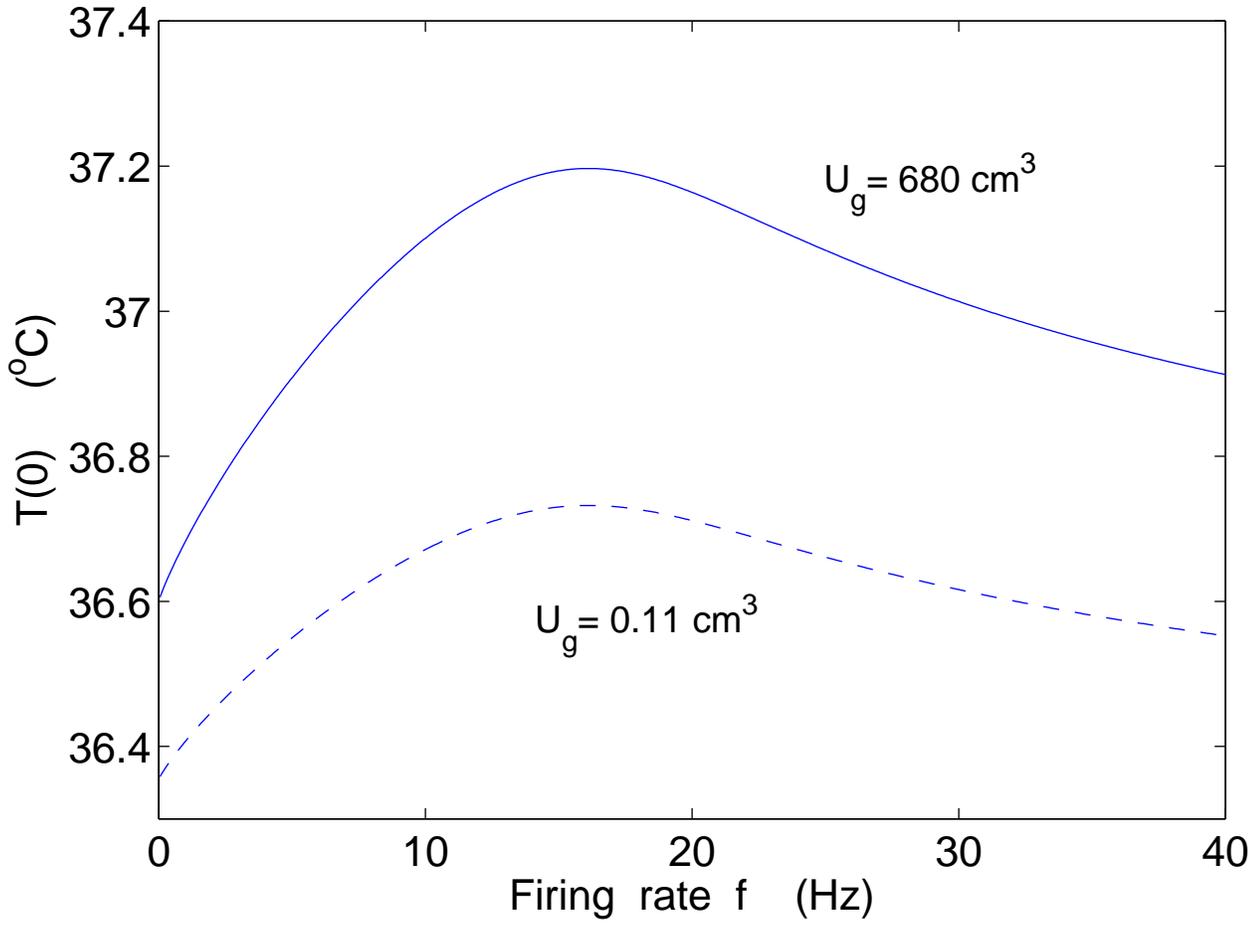
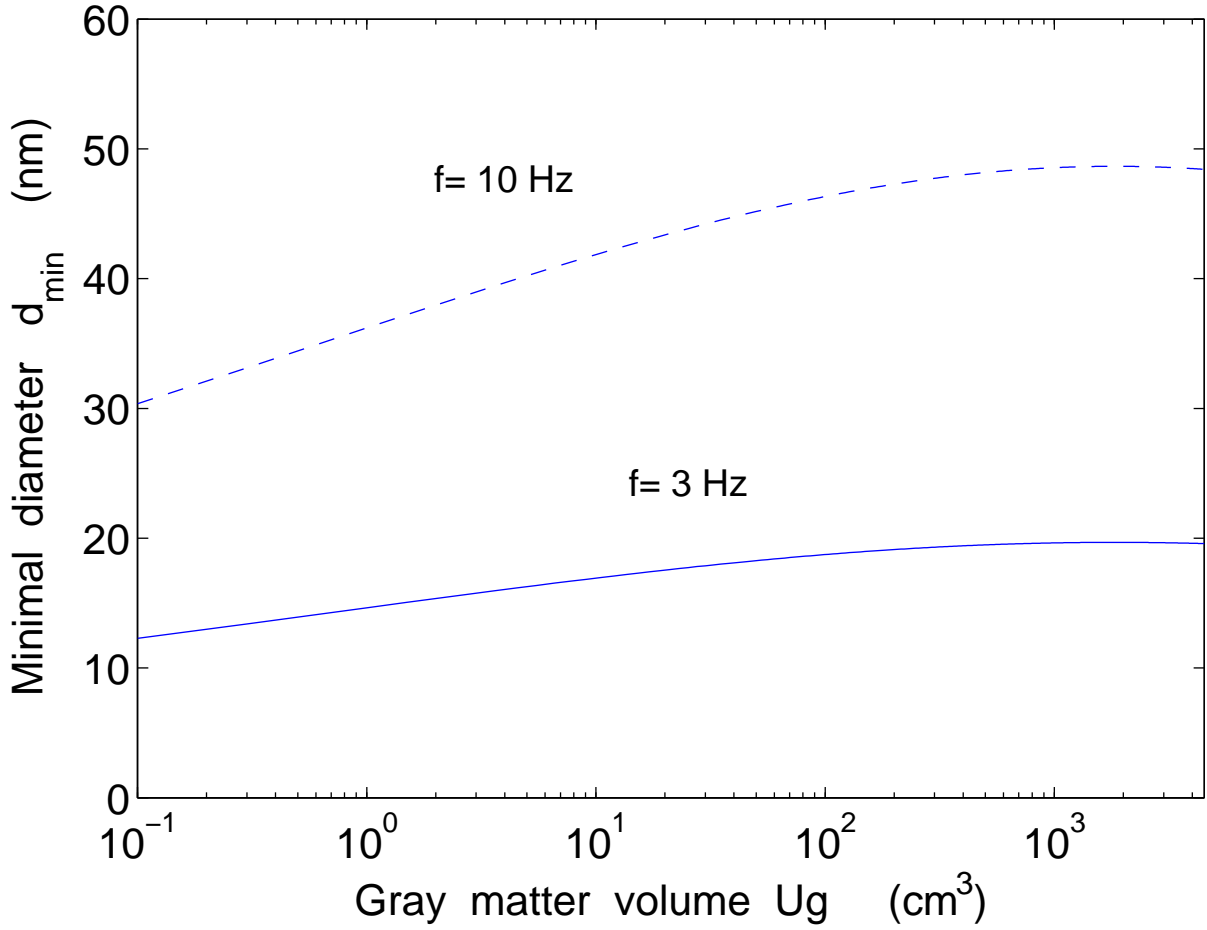
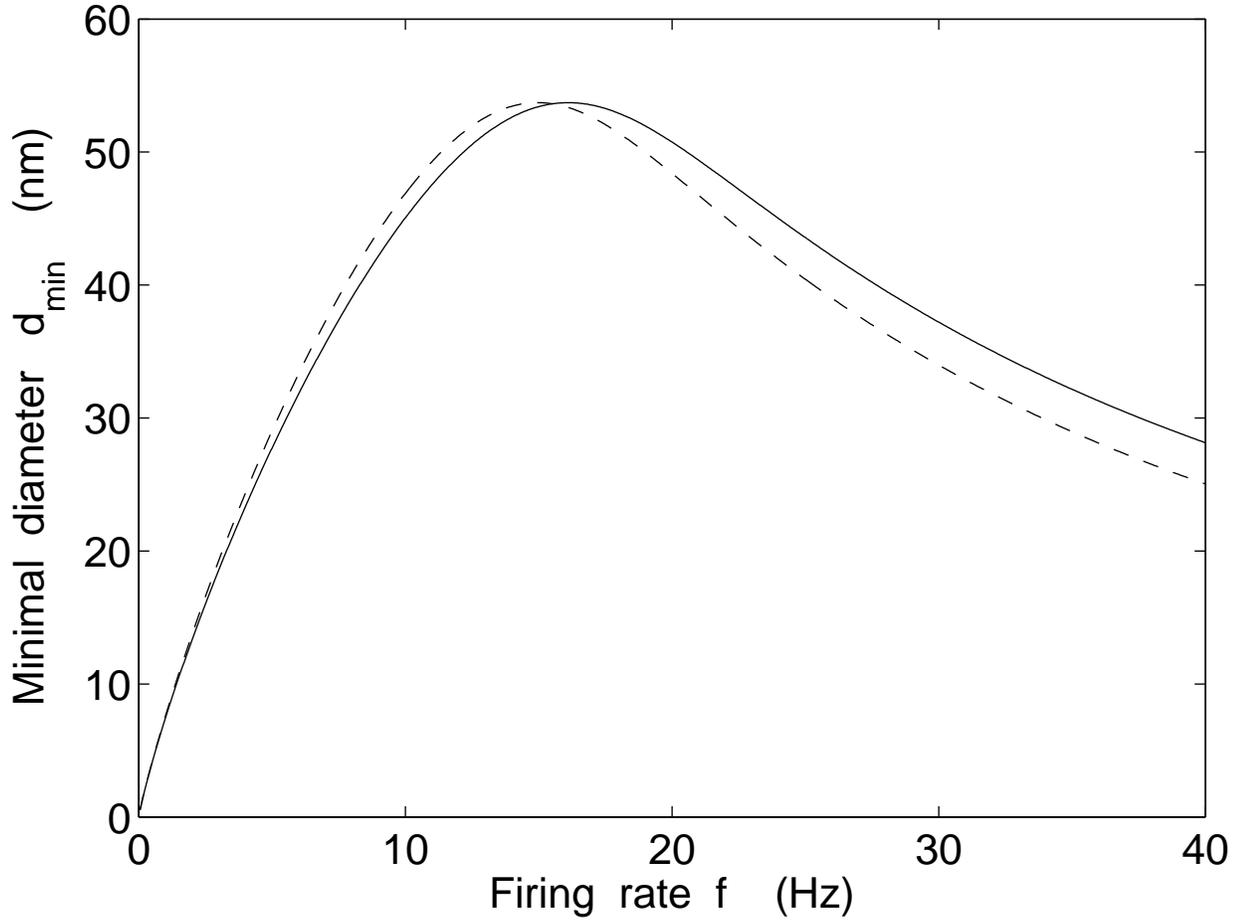
\end{document}